\documentclass[final,1p,times,authoryear]{elsarticle}

\usepackage{bm}
\usepackage{amsthm, amssymb}
\usepackage{mathtools, cuted}
\usepackage{color}
\usepackage{xcolor}
\usepackage{multicol}
\usepackage[hidelinks]{hyperref}
\usepackage{enumitem}
\usepackage{graphicx}
\usepackage{float}
\usepackage{subcaption} 
\usepackage[toc,page]{appendix} 

\newtheorem*{remark}{Remark} 

\usepackage{multicol}
\usepackage{parskip}
\newcommand\scalemath[2]{\scalebox{#1}{\mbox{\ensuremath{\displaystyle #2}}}}
\newcommand{\abs}[1]{\vert {#1} \vert}

\newcommand{\KI}{K_\mathrm{I}}
\newcommand{\KII}{K_\mathrm{II}}
\newcommand{\KIII}{K_\mathrm{III}}
\newcommand{\KIcirc}{\overline{K_\mathrm{I}}}
\newcommand{\KIpert}{\widetilde{K_\mathrm{I}}}
\newcommand{\KIIcirc}{\overline{K_\mathrm{II}}}
\newcommand{\KIIpert}{\widetilde{K_\mathrm{II}}}
\newcommand{\KIIIcirc}{\overline{K_\mathrm{III}}}
\newcommand{\KIIIpert}{\widetilde{K_\mathrm{III}}}
\newcommand{\Gcirc}{\overline{G}}
\newcommand{\Gpert}{\widetilde{G}}

\newcommand{\Gc}{G_\mathrm{c}}
\newcommand{\da}{\Delta{a}}
\newcommand{\fract}{\mathcal{L}}
\newcommand{\Hilbert}{\mathcal{H}}
\newcommand\norm[1]{\lVert#1\rVert}

\usepackage{lineno}
\journal{Computer Methods in Applied Mechanics and Engineering}

\begin{document}
	\title{Bridging perturbation and variational approaches in brittle fracture.}
	\author[add1]{Serafim Egorov}
        \author[add2]{Antoine Sanner}
        \author[add1]{Jean Sulem}
        \author[add3]{Lars Pastewka}
	\author[add1]{Mathias Lebihain}
	\cortext[cor1]{Corresponding author : serafim.egorov@enpc.fr}
	\address[add1]{Navier, Ecole des Ponts, Univ Gustave Eiffel, CNRS, Marne-la-Vall{\'e}e, FR}
	\address[add2]{Computational Mechanics of Building Materials, Institute for Building Materials, ETH Zürich, Zürich, CH}
        \address[add3]{Department of Microsystems Engineering (IMTEK), University of Freiburg, Freiburg, DE}
	
\begin{abstract}
    We present a variational reduced-order model for three-dimensional coplanar propagation of sharp cracks in heterogeneous perfectly brittle solids under mixed-mode I+{II}+{III} loading. The approach connects the variational fracture formulation of \cite{francfort_revisiting_1998} and the perturbation theory of \cite{rice_first-order_1985} by computing equilibrium crack-front configurations through minimization of the total energy defined as the sum of (i) the elastic potential energy, evaluated asymptotically from front deformations, and (ii) the dissipated energy, set by the fracture energy field.
    
    The potential energy and its derivatives are evaluated efficiently using the Fast Fourier Transform. The resulting nonconvex box-constrained minimization problem is solved with a matrix-free Newton conjugate gradient algorithm with a trust region and physics-based preconditioning, enforcing irreversibility while resolving energy barriers and long-range elastic interactions. We validate our implementation against newly derived analytical solutions. We then perform 116,000 large-scale simulations of tensile and shear crack  propagation in disordered media to quantify the impact of finite-size effects, disorder intensity, and mode mixity. The simulations reproduce the transition from smooth to intermittent crack growth, and show that mode mixity has limited influence on the onset of intermittency but induces quasi-elliptic fronts in mixed {II}+{III} loading. They reveal a size-dependent crossover from disorder-induced weakening to toughening controlled by the emergence of depinning instabilities.
\end{abstract}
		
	
\maketitle
	
\section{Introduction}
\label{sec}

Modeling 3D fracture in perfectly brittle materials is numerically challenging and computationally demanding due to the presence of singular stress concentrations localized in the vicinity of a propagating crack front. The presence of heterogeneities further complicates modeling, as crack growth unfolds as a highly intermittent process, alternating between pinning phases, whose durations span several time scales, and avalanches, whose sizes scale from the heterogeneity size to that of the system \citep{bonamy_failure_2011}. Addressing these challenges requires dedicated numerical methods that can describe the multiscale nature of fracture processes in heterogeneous media.

Alongside the extended finite element method (XFEM) \citep{moes_finite_1999} and cohesive zone modeling \citep{xu_numerical_1994}, the phase-field method \citep{bourdin_variational_2008}, which is built upon \cite{francfort_revisiting_1998}'s variational approach to fracture, has established itself as a robust technique to model 3D crack propagation in materials containing fluctuations of mechanical properties \citep{clayton_geometrically_2014, nguyen_large-scale_2017, henry_pinning_2024, schneider_fft-based_2025, aranda_crack-length_2025}. However, this method requires expensive volumetric meshing, and it does not keep the crack sharp, but introduces a length scale $\ell$ over which diffuse damage spreads. While this length scale is crucial to capture crack nucleation \citep{tanne_crack_2018}, it may interact with the heterogeneity size \citep{hossain_effective_2014, li_meso-scale_2024} and modify crack growth.

Independently, approaches based on \cite{rice_first-order_1985}'s perturbation theory have brought valuable physical insights on the failure of heterogeneous brittle materials, while keeping cracks sharp (see a recent review from \cite{lazarus_asymptotic_2025}). These approaches can be considered as a reduced-order model for brittle fracture, as 3D crack propagation is simulated through the meshing of the sole crack front. This dimensionality reduction comes with a significant drop in computational cost, allowing for the modeling of crack interaction with millions of inclusions in a matter of hours on a single core computer \citep{lebihain_effective_2021}. However, crack growth is often modeled using an \textit{ad hoc} viscous regularization of Griffith's criterion, which can affect the fracture process itself \citep{bares_fluctuations_2014}.  In addition, these studies are mainly focused on \emph{semi-infinite} cracks propagating in \emph{tensile} mode I \citep{gao_first-order_1989, patinet_finite_2013}. Yet, cracks are of finite size and sometimes can be loaded in mixed (tensile and shear) modes I+II+III (e.g. interface debonding in multilayered composites, sheared contacts, fault motion, etc.). As a result, the influence of heterogeneities, size effects, and mode mixity on three-dimensional cracks is not yet fully understood.

In this work, we propose a variational approach for coplanar quasi-static crack propagation in heterogeneous brittle media, bridging the formulation of \cite{francfort_revisiting_1998} and \cite{rice_first-order_1985}'s perturbation theory. Equilibrium positions of the crack front are computed by \emph{minimizing} the sum of the potential energy of the system that depends on loading and the crack geometry, and the dissipated energy determined by the (heterogeneous) fracture energy field. Unlike previous approaches that are based on the resolution of Griffith's criterion through an asymptotic expansion of the energy release rate, our energy-based method directly builds on the asymptotic estimate of its potential energy, which is efficiently computed using the Fast Fourier Transform (FFT). We further leverage methods from non-convex optimization to satisfy the physical constraints imposed by the fracture processes and seek further speed up with efficient preconditioning of the system.

Our paper is organized as follows. In Section~\ref{sec:theory}, we derive our variational perturbative approach for the coplanar crack growth under mixed mode I+II+III loading. This involves the rigorous calculation of the asymptotic expansion of the potential energy for an elastic body containing a single quasi-circular crack, inspired by the recent work of \cite{sanner_crack-front_2022} in contact mechanics. Our model is then implemented in Section~\ref{sec:numerics} and validated against newly-derived analytical benchmarks. Finally, we show in Section~\ref{sec:applications} some applications of our work related to: i) crack propagation in disordered medium; ii) intermittent energy dissipation; iii) and material toughening. These applications are supported by the results of 116,000 large-scale simulations of shear and tensile crack propagation in disordered media.

\section{Variational formulation of brittle fracture in a perturbative framework}
\label{sec:theory}

To build our method, we start from the variational approach to brittle fracture of \cite{francfort_revisiting_1998}, where the equilibrium configuration of the crack, represented by the set $\Gamma(t)$, comprising a two-dimensional crack surface and its one-dimensional boundary, the crack front, at time $t$, can be computed from the minimization:
\begin{equation}
\label{eq:variational_formulation}
    \min_{\Gamma^* \supseteq \Gamma(t'), t' \leq t} \Pi_\mathrm{pot}(\Gamma^*) + \Pi_\mathrm{dis}(\Gamma^*),
\end{equation}
where (i) $\Pi_\mathrm{pot}(\Gamma^*)$ is the \emph{potential energy} (elastic energy minus work of external forces) of the system constituted of the crack set $\Gamma^*$ and the embedding structure, and (ii) $\Pi_\mathrm{dis}(\Gamma^*)$ is the \emph{energy dissipated} during the entire history of the fracture process. $\Gamma(t')$ represents the crack configuration at a previous loading increment $t'$, and $\Gamma^* \supseteq \Gamma(t')$ denotes the set of admissible crack configurations at the current loading increment $t \geq t'$ that contains the previous crack set $\Gamma(t')$. This enforces the irreversibility of crack growth, ensuring that the crack set evolves monotonically (i.e. the crack cannot heal).

While $\Pi_\mathrm{dis}$ can be swiftly estimated from the integration of the fracture energy over the cracked surface, $\Pi_\mathrm{pot}$ is only known for very simple crack geometries (semi-infinite crack, penny-shaped internal crack, circular external connection, etc.; see \cite{tada_stress_2000}). In this section, we propose a systematic method to calculate $\Pi_\mathrm{pot}$ for geometries slightly deviating from a reference configuration, using Bueckner-Rice's weight function theory \citep{rice_weight_1989}.  Note that alternative variational formulations, also combined with perturbative methods, have been proposed in the context of quasi-static fracture by \cite{salvadori_minimum_2013} and \cite{salvadori_fracture_2016}.

\subsection{Differential equation for the potential energy of a quasi-circular crack}
\label{subsec:potential_energy}

\begin{figure}[h]
		\centering
		\noindent\includegraphics[width=0.45\textwidth]{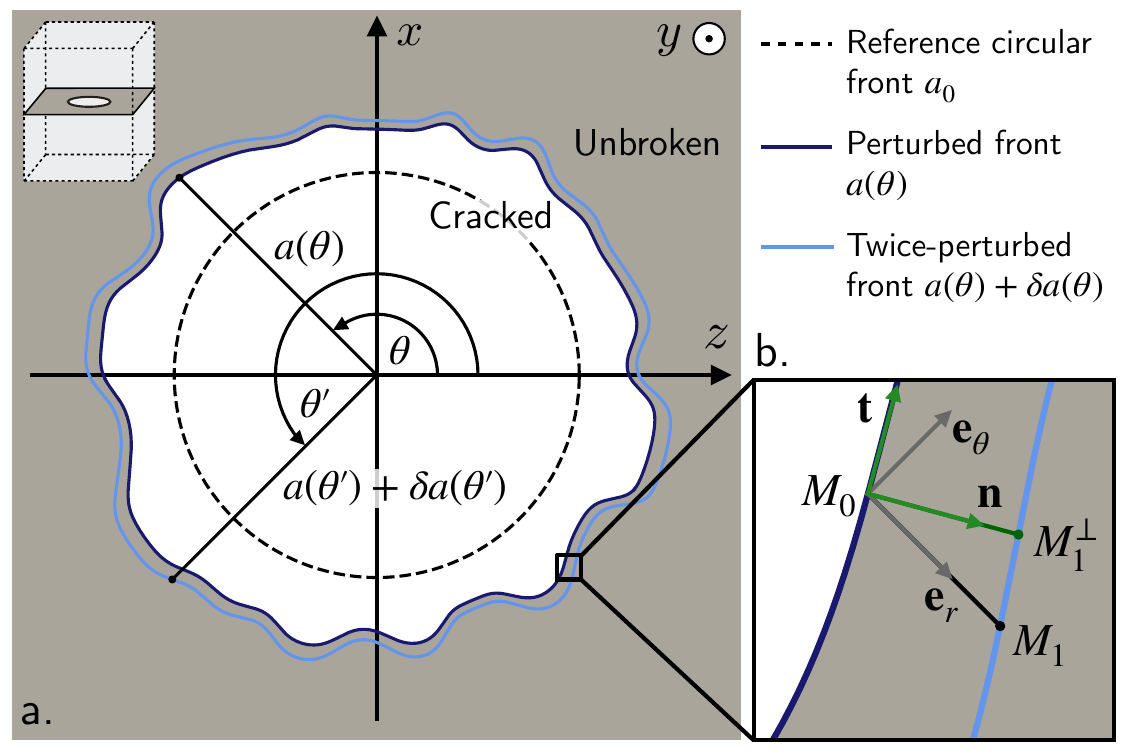}
		\caption{a. The reference penny-shaped crack with a circular front of radius $a_0$ (in dashed black line) is perturbed within its plane $(xOz)$ by a small amount $\Delta a(\theta) = A\phi(\theta)$ in the direction $\mathbf{e_r}$, giving rise to a perturbed crack front parametrized by $a(\theta) = a_0 + \Delta a(\theta)$ (in solid dark blue line). This already perturbed crack front is perturbed again by an infinitesimal quantity $\delta a(\theta) = \delta A\phi(\theta)$ in the direction $\mathbf{e_r}$ to form a twice-perturbed crack front (in solid light blue line). b. This second transformation can be equivalently applied either in direction $\mathbf{e_r}$ with amplitude $\delta a(\theta)$ ($M_0 \rightarrow M_1$) or in the direction $\mathbf{n}$ normal to the crack front with amplitude $\delta a_\mathrm{\perp}(\theta)$ ($M_0 \rightarrow M_1^\perp$).}
		\label{fig:perturbed_crack}
	\end{figure}

We consider a planar penny-shaped crack embedded in an infinite medium made of some isotropic and linearly elastic material of Young's modulus $E$ and Poisson's ratio $\nu$. The reference crack front $\mathcal{F}_{0}$ describes a circle of radius $a_{0}$, centered on a point $O$, chosen as the origin of our coordinate system (see the black dashed line in Fig.~\ref{fig:perturbed_crack}a). The crack is and remains contained in the $(xOz)$ plane. It is loaded in mixed mode I+{II}+{III} through some arbitrary system of shear and tensile forces that gives rise to unperturbed stress intensity factors $\KIcirc$, $\KIIcirc$ and $\KIIIcirc$ along the reference circular crack front $\mathcal{F}_{0}$. Then, the position of the crack front is perturbed \emph{within its plane} by a small quantity $\Delta a$ in the radial direction $\mathbf{e_r}$, giving rise to a new crack front $\mathcal{F}$ parameterized by $a(\theta) = a_0 + \Delta a(\theta)$ (see the dark blue solid line in Fig.~\ref{fig:perturbed_crack}a). $\Pi_\mathrm{pot}([a])$ denotes the potential energy of the system composed of the crack with perturbed front $\mathcal{F}$ and the embedding structure. The bracket notation $[a]$ indicates a functional dependence on the entire front configuration $a(\theta)$.

Our goal is to calculate $\Pi_\mathrm{pot}([a])$ asymptotically. This was recently achieved by \cite{sanner_crack-front_2022} in the case of a perturbed Johnson-Kendall-Roberts (JKR) contact by guessing the expression of $\Pi_\mathrm{pot}([a])$ that satisfies the variation $\delta\Pi_\mathrm{pot}/\delta a \approx + 2\pi a G$, where $G$ is the energy release rate along an arbitrarily perturbed crack front\footnote{Note that the plus sign comes from the fact that the circular contact edge propagates in the inward direction $-\mathbf{e_r}$}. Our objective here is to propose a \emph{generic method} to derive \emph{a closed-form formula} for the potential energy of a quasi-circular crack using \cite{rice_weight_1989}'s perturbation theory.

To do so, we will follow an idea used by \cite{leblond_second-order_2012} to derive the second-order expansion of the mode I stress intensity factor $\KI$ for the semi-infinite crack. The front perturbation $\Delta a(\theta)$ is decomposed as $\Delta a = A \phi(\theta)$, where $A$ is a small parameter that denotes the intensity of the perturbation and $\phi$ is a shape function. Our goal is to calculate the Gateaux derivative of the potential energy $\Pi_\mathrm{pot}$ at $a = a_0 + A \phi(\theta)$ in the direction $\phi$.

For this, we write $\Pi_\mathrm{pot}([a])$ as:
\begin{equation}
    \Pi_\mathrm{pot}([a = a_0 + A\phi]) = \pi(a_0, A, [\phi]),
\end{equation}
where $\pi$ is a real-valued function of the real numbers $a_0$ and $A$, and a functional of $\phi$. Following \cite{leblond_second-order_2012}, we apply to the perturbed front $\mathcal{F}$ another radial perturbation $\delta a(\theta) = \delta A \phi(\theta)$ ($\delta A \ll A$) in the direction $\mathbf{e_r}$, giving rise to a twice-perturbed front parameterized by $a(\theta) + \delta a(\theta)$ (see the light blue solid line in Fig.~\ref{fig:perturbed_crack}a). This secondary perturbation can be interpreted as the vector $M_0M_1$ in the direction $\mathbf{e_r}$, shown in Fig.~\ref{fig:perturbed_crack}b. The variation $\delta \Pi_\mathrm{pot}$ of the potential energy between the dark blue and light blue configurations of Fig.~\ref{fig:perturbed_crack}a writes as:
    \begin{equation}
        \label{eq:Pi_pot_radial_perturbation}
        \delta \Pi_\mathrm{pot} = \Pi_\mathrm{pot}([a + \delta a]) - \Pi_\mathrm{pot}([a]) = \dfrac{\partial \pi}{\partial A}(a_0, A, [\phi])\,\delta A + \mathcal{O}(\delta A^2).
    \end{equation}

Alternatively, one can go from the dark blue configuration of Fig.~\ref{fig:perturbed_crack}a to the light blue configuration using a transformation $\delta a_\perp$ applied in the direction $\mathbf{n}$ perpendicular to the crack front. This secondary perturbation can be interpreted as the vector $M_0M_1^\perp$, shown in Fig.~\ref{fig:perturbed_crack}b. By definition of the energy release rate, the variation in the potential energy can be expressed as \citep{rice_first-order_1985}: 
\begin{equation}
    \label{eq:Pi_pot_normal_perturbation}
    \delta \Pi_\mathrm{pot} = - \int_{0}^{2\pi} G([a]; \theta) \delta{a}_{\perp}(\theta) \,\mathrm{d}s(\theta) + \mathcal{O}(\norm{\delta a_\perp}^2),
\end{equation}
where $G([a]; \theta)$ is the energy release rate and $\mathrm{d}s(\theta) = \sqrt{a(\theta)^2+a'(\theta)^2} \,d\theta $ the arclength along the perturbed front $\mathcal{F}$. Using geometric arguments based on Fig.~\ref{fig:perturbed_crack}b, one can prove that:
\begin{equation}
    \label{eq:from_ds_to_dtheta}
    \delta{a}_{\perp}(\theta) \mathrm{d}s(\theta) = \delta A \phi(\theta) a(\theta) \mathrm{d}\theta + \mathcal{O}(\delta A^2).
\end{equation}

Equations~\eqref{eq:Pi_pot_radial_perturbation} and \eqref{eq:Pi_pot_normal_perturbation} represent the same energy variation, since it corresponds to that associated with the same geometric transformation (from the dark blue line to the light blue line in Fig.~\ref{fig:perturbed_crack}a). Using the uniqueness of Taylor's expansion for $\pi$, one finds:
\begin{equation}
        \label{eq:Pi_pot_differential equation}
    \dfrac{\partial \pi}{\partial A}(a_0, A, [\phi]) = - \int_{0}^{2\pi} G([a]; \theta) \phi(\theta) a(\theta) \,\mathrm{d}\theta.
\end{equation}

Equation~\eqref{eq:Pi_pot_differential equation} relates the variation in potential energy to the energy release rate $G([a]; \theta)$ of the perturbed configuration $a(\theta)$. Note that we only considered expansion in terms of $\delta A$ so that Eq.~\eqref{eq:Pi_pot_differential equation} stands at all orders in $A$. This means that any asymptotic expansion of $G([a]; \theta) = g(a_0, A, [\phi]; \theta) = g^\mathbf{0}(a_0, \theta) + g^\mathbf{1}(a_0, A, [\phi]; \theta) + ...$ at order $N$ in $A$ results, by integration of Eq.~\eqref{eq:Pi_pot_differential equation} in an asymptotic expansion of $\Pi_\mathrm{pot}([a])$ at order $N+1$ in $A$. The former is given by \cite{rice_first-order_1985}'s perturbation theory, which provides asymptotic estimates of the stress intensity factors (SIFs) $(K_p)_{p\in\{\mathrm{I,II,III}\}}$ and thus of energy release rate using \cite{irwin_analysis_1957}'s formula.
    
\subsection{Second-order expansion of the potential energy from \cite{gao_nearly_1988}'s linear theory}

In general, the SIFs along a crack front $a(\theta) = a_0 + \da(\theta)$ perturbed from its reference circular configuration of radius $a_0$ can be expressed as \citep{favier_coplanar_2006}:
\begin{equation}
\label{eq:SIFs}
    \scalemath{0.9}{
        \begin{aligned}    
    & \KI([a]; \theta) = \KIcirc(a_0, \theta) + \KIpert(a_0, [\KIcirc], [\da]; \theta) + \mathcal{O}(\norm{\Delta{a}}^2), \\
    & \KII([a]; \theta) = \KIIcirc(a_0, \theta) + \KIIpert(a_0, [\KIIcirc], [\KIIIcirc], [\da]; \theta) + \mathcal{O}(\norm{\Delta{a}}^2), \\
    & \KIII([a]; \theta) = \KIIIcirc(a_0, \theta) + \KIIIpert(a_0, [\KIIcirc], [\KIIIcirc], [\da]; \theta) + \mathcal{O}(\norm{\Delta{a}}^2),
    \end{aligned}
    }
\end{equation}
where $\KIcirc$, $\KIIcirc$, $\KIIIcirc$ are the stress intensity factors along the reference circular front and $\KIpert$, $\KIIpert$, $\KIIIpert$ represent their perturbation arising from the front roughness $\da(\theta)$. In the following, we use the decomposition of a $2\pi$-periodic function $f$ in Fourier series:
\begin{equation}
    f(\theta) = \sum_{k=-\infty}^{+\infty} \hat{f}_ke^{ik\theta} \Leftrightarrow \hat{f}_k=\dfrac{1}{2\pi} \int_{0}^{2\pi} f(\theta)e^{-ik\theta}\,\mathrm{d}\theta,
\end{equation}
where $\hat{f}_k$ is the $k$-th coefficient of the Fourier series of $f$. In particular, $\hat{f}_0$, or equivalently $(\widehat{f\,})_0$, denotes the $0$-th Fourier coefficient, i.e., the average value of $f$.

    The first order perturbation terms for SIFs have been obtained by \cite{gao_somewhat_1987} for mode I and \cite{gao_nearly_1988} for modes {II+III}. They read: 
\begin{equation}
\label{eq:K1_perturbed}
    \scalemath{0.9}{
    \begin{aligned}
    \KIpert(a_0, [\KIcirc], [\da]; \theta) & = \frac{\partial{\KIcirc}}{\partial{a}}(a_0, \theta)\Delta{a}(\theta) \\
    & - \frac{1}{2a_0}\left(\fract[\KIcirc\Delta{a}](\theta) - \fract[\KIcirc](\theta)\Delta{a}(\theta) \right) + \mathcal{O}(\norm{\Delta{a}}^2),
    \end{aligned}
    }
\end{equation}
and:
\begin{equation}
\label{eq:K2_perturbed}
    \scalemath{0.9}{
        \begin{aligned}
    \KIIpert(a_0, [\KIIcirc], [\KIIIcirc], [\da]; \theta) & = \frac{\partial{\KIIcirc}}{\partial{a}}(a_0, \theta)\Delta{a}(\theta) \\
    & - \frac{2-3\nu}{2(2-\nu)a_0}\left(\fract[\KIIcirc\Delta{a}](\theta) - \fract[\KIIcirc](\theta)\Delta{a}(\theta) \right) \\ 
    & - \frac{1}{(2-\nu)a_0}\left((\widehat{\KIIcirc\Delta{a}})_0-(\widehat{\KIIcirc})_0\Delta{a}(\theta)) \right) \\
    & - \frac{2}{(2-\nu)a_0}\KIIIcirc(a_0, \theta) \Delta{a}'(\theta) \\ 
    & + \frac{1}{(2-\nu)a_0}\left((\Hilbert[\KIIIcirc\Delta{a}](\theta) - \Hilbert[\KIIIcirc](\theta)\Delta{a}(\theta) \right) + \mathcal{O}(\norm{\Delta{a}}^2),
    \end{aligned}
    }
\end{equation}
and :
\begin{equation}
\label{eq:K3_perturbed}
    \scalemath{0.9}{
        \begin{aligned}
    \KIIIpert(a_0, [\KIIIcirc], [\KIIcirc], [\da]; \theta) &  = \frac{\partial{\KIIIcirc}}{\partial{a}}(a_0, \theta)\Delta{a}(\theta) \\
    & - \frac{2+\nu}{2(2-\nu)a_0}\left(\fract[\KIIIcirc(a_0)\Delta{a}](\theta) - \fract[\KIIIcirc(a_0)](\theta)\Delta{a}(\theta) \right) \\ 
    & - \frac{1-\nu}{(2-\nu)a_0} \left((\widehat{\KIIIcirc(a_0)\Delta{a}})_0-(\widehat{\KIIIcirc(a_0)})_0\Delta{a}(\theta) \right) \\
    & + \frac{2(1-\nu)}{(2-\nu)a_0}\KIIcirc(a_0, \theta) \Delta{a}'(\theta) \\ 
    & - \frac{1-\nu}{(2-\nu)a_0} \left(\Hilbert[\KIIcirc(a_0)\Delta{a}](\theta) - \Hilbert[\KIIcirc(a_0)](\theta)\Delta{a}(\theta) \right)+ \mathcal{O}(\norm{\Delta{a}}^2),
    \end{aligned}
    }
\end{equation}
where in Eqs.~(\ref{eq:K1_perturbed}-\ref{eq:K3_perturbed}), the implicit dependence of the reference SIFs $\overline{K_p}$ in $a_0$ and $\theta$ has sometimes been omitted for the sake of simplicity. $\fract[f]$ is the square-root (fractional) Laplacian and $\Hilbert[f]$ is the Hilbert transform of a $2\pi$-periodic function $f$, defined accordingly to the literature \citep{caffarelli_extension_2007, duoandikoetxea_fourier_2000}:
\begin{equation}
\label{eq:Operator_L}
    \fract[f](\theta) = \sum_{k = - \infty}^{+\infty
    } +|k|\hat{f_{k}}e^{\mathbf{i}k\theta} = \dfrac{1}{4\pi} \mathrm{PV}\int_{0}^{2\pi} \dfrac{f(\theta')-f(\theta)}{\sin^2\left[(\theta'-\theta)/2\right]} \,\mathrm{d}\theta',
\end{equation}
and: 
\begin{equation}
\label{eq:Operator_S}
    \Hilbert[f](\theta) = \sum_{k = - \infty}^{+\infty
    } -\mathbf{i}sgn(k)\hat{f_{k}}e^{\mathbf{i}k\theta} = \dfrac{1}{2\pi} \mathrm{PV}\int_{0}^{2\pi} \dfrac{\cos\left[(\theta'-\theta)/2\right]}{\sin\left[(\theta'-\theta)/2\right]} \left(f(\theta')-f(\theta)\right) \,\mathrm{d}\theta',
\end{equation}
where $\mathrm{PV}$ stands for the Cauchy principal value of the integral. The important observation is that $\fract$ and $\Hilbert$ are non-local in the sense that $\fract[\Delta a](\theta)$ and $\Hilbert[\Delta a](\theta)$ involve the entire roughness $\Delta a$ of the crack front $\mathcal{F}$, and not its sole value or derivatives at $\theta$. This corresponds to long-range interactions emerging from a linear elastic behavior \citep{tanguy_weak_2004}.
    
    \begin{remark}
    In this work, we follow the convention commonly used in partial differential equations and harmonic analysis, where \( \Delta = \partial_{xx} \) is negative semi-definite. Under this convention, the square-root (fractional) Laplacian \( (-\Delta)^s \) has a positive Fourier symbol, namely \( \widehat{(-\Delta)^s f}(k) = |k|^{2s} \hat{f}(k) \), and $\mathcal{L} = (-\Delta)^{1/2}$. For the Hilbert transform, we use the classical definition on the unit circle via its Fourier multiplier representation, which is consistent with standard harmonic analysis, though other sign conventions exist in the literature.
    \end{remark}
    
    The decomposition of the energy release rate $G$ follows the same as in Eq. ~\eqref{eq:SIFs} : 
\begin{equation}
    \label{eq:ERR}
    \scalemath{0.9}{
        \begin{aligned}    
    G(a_{0}, [a]; \theta) = \Gcirc(a_0, [\KIcirc], [\KIIcirc], [\KIIIcirc]; \theta) + \Gpert(a_{0}, [\KIcirc], [\KIIcirc], [\KIIIcirc], [\da]; \theta) + \mathcal{O}(\norm{\da}^2),
    \end{aligned}
    }
\end{equation}
where $\Gcirc$ is the zero-order term that corresponds to the circular crack of radius $a_0$ and $\Gpert$ is a first-order perturbation of $\Gcirc$ in $\da$.
    
Knowing the SIFs for the perturbed crack front $\mathcal{F}$ at first order in $\Delta{a}$ from Eqs.~\eqref{eq:K1_perturbed} for the tensile crack and from Eqs. ~\eqref{eq:K2_perturbed} and \eqref{eq:K3_perturbed} for shear, we can calculate the first-order perturbation of the energy release rate $\Gpert(a_{0}, [\KIcirc], [\KIIcirc], [\KIIIcirc], [\da]; \theta)$ using Irwin's formula:
    \begin{equation}
    \label{eq:Irwin_formula}
    G([a]; \theta) = \frac{1-\nu^2}{E} \left(\KI^2([a]; \theta) + \KII^2([a]; \theta)\right) + \frac{1+\nu}{E} \KIII^2([a]; \theta).
\end{equation}

$\Gpert$ can be expressed as: 
\begin{equation}
    \label{eq:G_perturbed}
    \begin{aligned}
    \Gpert(a_{0}, [\KIcirc], [\KIIcirc], [\KIIIcirc], [\da]; \theta) 
    & = \frac{2 (1-\nu^2)}{E} \KIcirc(a_0, \theta) \KIpert(a_0, [\KIcirc], [\da]; \theta)\\
    & + \frac{2 (1-\nu^2)}{E} \KIIcirc(a_0, \theta) \KIIpert(a_0, [\KIIcirc], [\KIIIcirc], [\da]; \theta)\\
    & + \frac{2 (1+\nu)}{E} \KIIIcirc(a_0, \theta)  \KIIIpert(a_0, [\KIIIcirc], [\KIIcirc], [\da]; \theta) + \mathcal{O}(\norm{\Delta{a}}^2).\\
    \end{aligned}
\end{equation}

The first-order expansion of $G$ in Eqs.~\eqref{eq:ERR} and \eqref{eq:G_perturbed} is injected into Eq.~\eqref{eq:Pi_pot_differential equation} to estimate $\Pi_\mathrm{pot}$ through simple integration with respect to the perturbation parameter $A$. We obtain the asymptotic expression for the potential energy at second order in $\Delta{a}$:
\begin{equation}
\label{eq:Pi_pot_perturbed}
    \scalemath{0.9}{
    \begin{aligned}
    \Pi_{\mathrm{pot}}(a_0, [\Delta a]) &= \Pi_{\mathrm{pot}}^{\mathrm{uncracked}} + \overline{\Pi}_{\mathrm{pot}}(a_0)  \\
    & - 2\pi a_0 (\widehat{\Gcirc\Delta{a}})_0 - \pi (\widehat{\Gcirc\Delta{a}^2})_0 - \pi a_{0} (\widehat{\frac{\partial{\Gcirc}}{\partial{a}}\Delta{a}^2})_0 \\
    & + \frac{2\pi}{E} \frac{(1-\nu^2)}{2}\KIcirc^2\widehat{(\fract[\Delta{a}]\Delta{a}})_0\\ 
    & + \frac{2\pi}{E} \frac{(2-3\nu)(1-\nu^2)}{2(2-\nu)} \Bigl(\widehat{(\KIIcirc\Delta{a}\fract[\KIIcirc\Delta{a}]})_0 - \widehat{(\KIIcirc\fract[\KIIcirc]\Delta{a}^2)}_0\Bigr) \\ 
    & + \frac{2\pi}{E} \frac{(2+\nu)(1+\nu)}{2(2-\nu)} \Bigl(\widehat{(\KIIIcirc\Delta{a}\fract[\KIIIcirc\Delta{a}]})_0 - \widehat{(\KIIIcirc\fract[\KIIIcirc]\Delta{a}^2)}_0\Bigr) \\
    & + \frac{2\pi}{E}\frac{(1-\nu^2)}{(2-\nu)}\Bigl((\widehat{\KIIcirc\Delta{a}})_0^2 + (\widehat{\KIIIcirc\Delta{a}})_0^2 - (\widehat{\KIIcirc})_0 (\widehat{\KIIcirc\Delta{a}^2})_0 - (\widehat{\KIIIcirc})_0 (\widehat{\KIIIcirc\Delta{a}^2})_0\Bigr) \\
    & - \frac{2\pi}{E}\frac{(1-\nu^2)}{(2-\nu)}\Bigl(\widehat{(\KIIcirc\Delta{a}\Hilbert[\KIIIcirc\Delta{a}]})_0 - \widehat{(\KIIcirc\Hilbert[\KIIIcirc]\Delta{a}^2)}_0\Bigr) \\
    & + \frac{2\pi}{E}\frac{(1-\nu^2)}{(2-\nu)}\Bigl(\widehat{(\KIIIcirc\Delta{a}\Hilbert[\KIIcirc\Delta{a}]})_0 - \widehat{(\KIIIcirc\Hilbert[\KIIcirc]\Delta{a}^2)}_0\Bigr)+ \mathcal{O}(\norm{\Delta{a}}^3),
    \end{aligned}
    }
\end{equation}
where $\Pi_{\mathrm{pot}}^{\mathrm{uncracked}}$ is the potential energy of the system in the absence of a crack, and $\overline{\Pi}_{\mathrm{pot}}$ is the zero-order term that corresponds to the potential energy of the circular crack of radius $a_0$. We set $\Pi_{\mathrm{pot}}^{\mathrm{uncracked}}$ to zero, since it does not impact the minimization.

The missing ingredient to complete the variational formulation of Eq.~\eqref{eq:variational_formulation} is now the dissipated energy, which independently of the loading mode can be expressed as:
\begin{equation}
    \label{eq:Pi_dis}
    \Pi_\mathrm{dis}([a]) = \int_{0}^{2\pi}\int_{0}^{a(\theta)} G_\mathrm{c}(r, \theta) \, r \mathrm{d}r \,\mathrm{d}\theta,
\end{equation}
where $G_\mathrm{c}$ is the fracture energy along the $(xOz)$ plane.

As a result, by combining Eqs.~\eqref{eq:Pi_pot_perturbed} and \eqref{eq:Pi_dis}, which depend solely on the crack front position $a(\theta)$, together with the variational formulation of Eq.~\eqref{eq:variational_formulation} allows to calculate the equilibrium positions of $a(\theta)$ by solving an associated minimization problem. Note that, on the parts of the crack front where the irreversibility constraint is inactive, the stationarity of the total energy $\Pi_\mathrm{tot} = \Pi_\mathrm{pot} + \Pi_\mathrm{dis}$ gives:
\begin{equation}
\label{eq:stationarity_conditions}
    \dfrac{\delta\Pi_\mathrm{tot}}{\delta a}([a]; \theta) = \dfrac{\delta\Pi_\mathrm{pot}}{\delta a}([a]; \theta) + \dfrac{\delta\Pi_\mathrm{dis}}{\delta a}([a]; \theta)=0,
\end{equation}
and one can prove from Eqs.~\eqref{eq:G_perturbed}, \eqref{eq:Pi_pot_perturbed} and \eqref{eq:Pi_dis} that:
\begin{equation*}
    \dfrac{\delta\Pi_\mathrm{pot}}{\delta a}([a]; \theta) = -2\pi a(\theta) G([a]; \theta) + \mathcal{O}(\norm{\Delta a}^2) \text{, and } \dfrac{\delta\Pi_\mathrm{dis}}{\delta a}([a]; \theta) = +2 \pi a(\theta) G_\mathrm{c}(r=a(\theta),\theta).
\end{equation*}
so that we retrieve Griffith's criterion. On the parts of the crack front where the irreversibility constraint is active, the crack does not grow and $G < \Gc$.

\section{Numerical implementation}
\label{sec:numerics}

\cite{rice_first-order_1985}'s perturbation theory is traditionally used to model a 3D crack propagation through a viscous regularization of Griffith's criterion, based on a rate-dependent fracture energy \citep{gao_first-order_1989, lebihain_towards_2021} or a fatigue Paris law \citep{lazarus_numerics_2003, vasoya_fingering_2016}. When studying crack propagation in the quasistatic limit, it may introduce numerical artifacts that change the front deformation \citep{favier_statistics_2006} and its dynamics \citep{bares_fluctuations_2014, ponson_crack_2017} due to the presence of finite viscosity in the model. Our goal here is to simulate crack growth in a \textit{perfectly brittle material}, using the perturbative variational framework introduced in Eqs.~\eqref{eq:variational_formulation}, \eqref{eq:Pi_pot_perturbed} and \eqref{eq:Pi_dis}. We show that appropriate solver choices for the associated non-convex optimization problem permits (i) to satisfy the physics of crack growth in heterogeneous media, and (ii) accelerate computations through physics-based preconditioning. The proposed implementation is finally validated against novel analytical solutions, obtained using the perturbation theory.

\subsection{Constraints on the minimization algorithm}
\label{subsec:constraints}

In the following, the crack front is discretized in $N$ points $a_i = a(\theta_i)$, ${i\in [0,N-1]}$, where $\theta_i = 2\pi/N \times i$. Solving Equation~\eqref{eq:variational_formulation} consists in finding the $a_i$ that minimize the total energy $\Pi_\mathrm{tot} = \Pi_\mathrm{pot} + \Pi_\mathrm{dis}$ at fixed load, subject to irreversibility. As shown below, spatial fluctuations of $\Gc$ can make $\Pi_\mathrm{tot}$ nonconvex, even when the homogeneous reference is stable ($\partial \Gcirc/\partial a < 0$), so its minimization requires dedicated non-convex optimization tools. Moreover, the physics of crack growth impose additional requirements that must be encoded as numerical constraints and embedded in the solver architecture.

\subsubsection{Box constraints to satisfy crack irreversibility}

Crack propagation in brittle media is an irreversible process. It implies that the front position $(a_i)_{i\in [0,N-1]}$ at the current step must be greater than that at the previous step, $a_i \geq a_i^\mathrm{prev}$, $\forall\, i\in [0,N-1]$. As a result, our minimization solver should be able to handle \textit{bound constraints}.
    
\begin{figure}[h]
		\centering
		\noindent\includegraphics[width=0.95\textwidth]{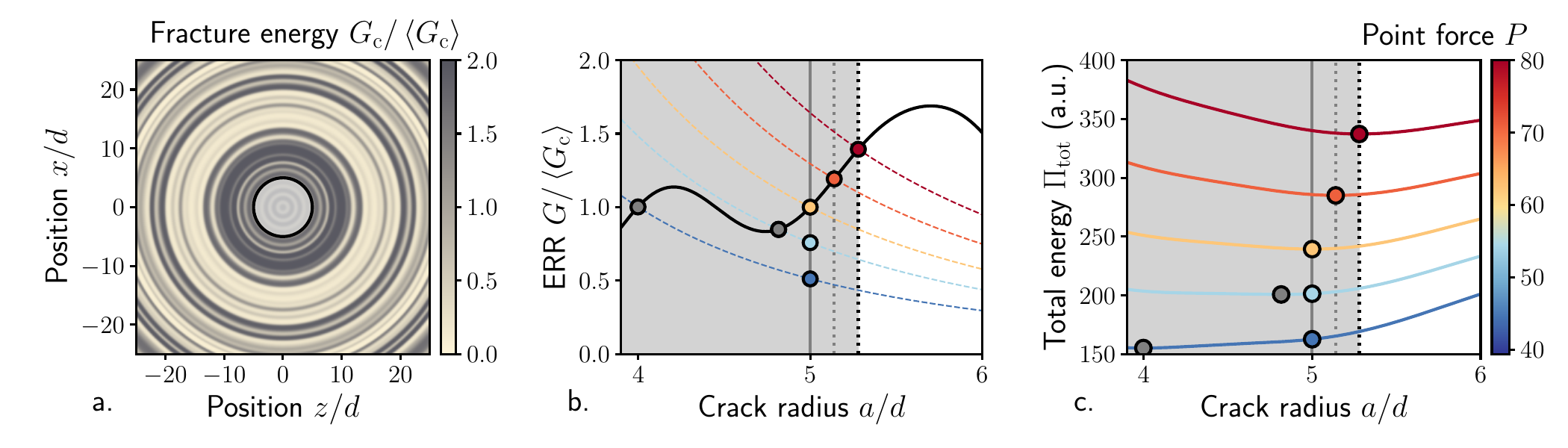}
		\caption{a. A mode I crack of initial radius $a_\mathrm{ini}$ (solid black line) propagates within its plane $(xOz)$ under a pair of symmetric normal forces $P$ applied at the center $O$. The fracture energy $\Gc$ varies with the radial coordinate $r$ over a characteristic length scale $d$. b. Griffith’s criterion: the crack propagates when the energy release rate $G(P,a)$ (thin colored dashed lines, from blue to red with increasing $P$) equals the local fracture energy $\Gc(a)$ (solid black line), provided the irreversibility condition $a \geq a^\mathrm{prev}$ is satisfied (solid grey vertical line for $a_\mathrm{ini} = 5\,d$, dotted grey/black lines thereafter). c. Total energy $\Pi_\mathrm{tot}$ as a function of crack radius $a$. Equilibrium configurations satisfying Eq.~\eqref{eq:variational_formulation} are indicated by colored circles. Grey circles show unconstrained minimizers of $\Pi_\mathrm{tot}$ (i.e., without enforcing irreversibility).}
		\label{fig:bound_constraints}
	\end{figure}

Irreversibility is illustrated in Fig.~\ref{fig:bound_constraints}, with a ``simple'' example of the propagation of a circular crack in a heterogeneous fracture energy field under increasing tensile loading. The crack is initially circular with radius $a_\mathrm{ini}$. The loading consists of a pair of normal forces $P$ applied at the center $O$ of the initial crack. The fracture energy is a function of $r$ only, fluctuating around an average value $\left<G_\mathrm{c}\right>$ at a characteristic length scale $d$. As both the loading and the fracture energy are axisymmetric with respect to $O$, the crack remains circular during its propagation, and its evolution can be lumped into a single scalar value, the crack radius $a$. 

For small $P$, the energy release rate satisfies $G(P;a_\mathrm{ini}) < \Gc(a_\mathrm{ini})$, so the crack does not grow and $a=a_\mathrm{ini}$ (blue curves/markers in Fig.~\ref{fig:bound_constraints}b). Note that $G=\Gc$ might be met for a smaller radius (grey markers), but irreversibility prevents healing, i.e. $a$ cannot decrease below $a_\mathrm{ini}$. In energetic terms (Fig.~\ref{fig:bound_constraints}c), the admissible solution remains on the bound $a = a_\mathrm{ini}$ (blue marker), even if the unconstrained minimum of $\Pi_\mathrm{tot}$ lies at a smaller $a$ (grey marker). As $P$ increases, $G(P;a)$ grows until $G(P;a_\mathrm{ini})=\Gc(a_\mathrm{ini})$, at which point the crack advances (yellow to red curves/markers in Fig.~\ref{fig:bound_constraints}b–c). From that point, the evolution follows the minimizer of $\Pi_\mathrm{tot}$. Note that Fig.~\ref{fig:bound_constraints} clearly illustrates the equivalence between Griffith's criterion and our perturbative variational formulation.

\subsubsection{Trust Region constraints to ensure Griffith criterion}

During crack propagation in a heterogeneous medium, the crack may locally cross a region of sharp decrease in fracture energy, and become unstable \citep{nguyen_stability_2000, roux_effective_2003}. Considering negative geometries ($\partial \Gcirc/\partial a < 0$), the energy release rate will decrease during crack growth at fixed $P$ and the crack will arrest when it hits a barrier of fracture energy larger than its energy release rate. When $\Gc$ exhibits large fluctuations or in case of weak stabilization ($\partial \Gcirc/\partial a \simeq 0$), the total energy $\Pi_\mathrm{tot}$ may have multiple local minima. As the crack cannot cross regions where $G < \Gc$, physically-meaningful simulations must therefore select the \emph{next} admissible metastable minimum, rather than an arbitrary (even global) minimum. We enforce this via a trust-region method whose radius is set by the characteristic fluctuation scale $d$ of $\Gc$, preventing unphysical jumps and guiding the solver to the physical minimum.

\begin{figure}[h]
\centering
	\noindent\includegraphics[width=0.95\textwidth]{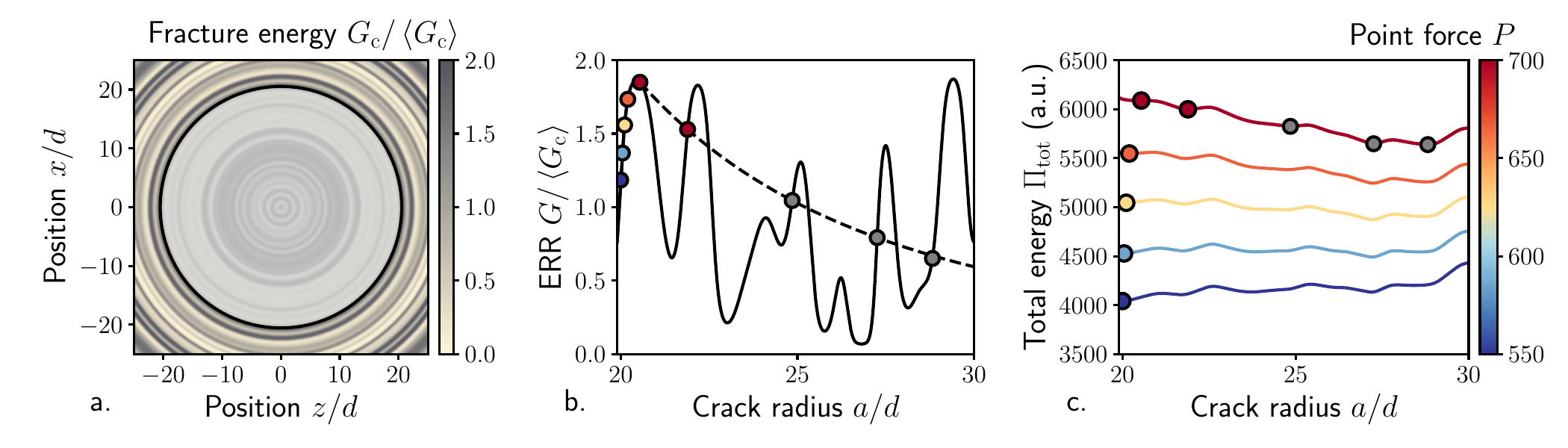}
	\caption{a. Crack propagation in a heterogeneous medium, following the configuration of Figure~\ref{fig:bound_constraints}. b. Under increasing load $P$, the crack advances through a sequence of stable equilibrium positions (blue to orange circles). When a local drop in fracture energy (solid black line) causes instability, the crack abruptly snaps to a new configuration (red circles), controlled by the decrease of $G$ at fixed load (dashed black line). Additional stable solutions to Griffith’s criterion $G(P;a) = \Gc(a)$ are shown as grey circles. c. Total energy $\Pi_\mathrm{tot}$ as a function of crack radius $a$. Equilibrium configurations satisfying Eq.~\eqref{eq:variational_formulation} are indicated by colored circles. Grey markers correspond to nonphysical local minima of $\Pi_\mathrm{tot}$, where the crack would propagate through energy barriers.
}
	\label{fig:trust_region}
\end{figure}

This behavior is illustrated in Fig.~\ref{fig:trust_region}, which shows an example of unstable crack propagation. For small $P$ (blue–orange markers in Fig.~\ref{fig:trust_region}b), the crack grows stably through a zone where $\partial \Gc/\partial r > 0$. Upon entering a region where $\Gc$ decreases, instability arises when $\partial G/\partial a > \partial \Gc/\partial r$ (red markers). At that stage, the equation $G=\Gc$ (intersection of the dashed $G$-curve with the solid $\Gc$-curve) admits multiple roots. The only admissible state is the \emph{first} stable equilibrium, satisfying $G=\Gc$ and the local stability condition $\partial G/\partial a \le \partial \Gc/\partial r$. In energetic terms (Fig.~\ref{fig:trust_region}c), the early solutions (blue–orange markers) lie in a region of $\Pi_\mathrm{tot}$ with positive curvature. The unstable configuration corresponds to a saddle point  (first red marker), and the only physically admissible state is the \emph{next} local minimizer of $\Pi_\mathrm{tot}$ (second red marker). Importantly, the global minimum of $\Pi_\mathrm{tot}$ is not a physical solution, so that the solution of Eq.~\eqref{eq:variational_formulation} can be a local minimizer. The role of the trust region is to restrict the allowable increment in the crack front position within each iteration to a bounded neighborhood of the current configuration, thereby preventing the algorithm from "jumping" through energy barriers. Note that, in standard trust-region methods the radius is increased whenever the quadratic model provides an accurate prediction of the objective decrease. In our setting, however, it is also necessary to enforce an \emph{upper bound} on the trust-region size. We set this maximum radius to a fraction of the characteristic heterogeneity length scale $d$ (of order a few tenths of $d$), because fluctuations of the total energy are primarily controlled by the heterogeneous fracture-energy field on that scale (see Fig.~\ref{fig:trust_region}c).

\subsubsection{Matrix-Free Algorithm with preconditioning}

Newton-type methods are well suited to our problem because they combine bound constraints and trust region, while exploiting second-order curvature information to achieve fast local convergence. We therefore adopt a truncated Newton-conjugate gradient (CG) scheme with trust region: at each outer iteration, we compute an inexact Newton step on the working set (i.e. the subset of variables that do not remain on the bounds during the step), by approximately solving the linearized system with conjugate gradients, and the trust region controls the step length to avoid skipping energy barriers. As $\Pi_\mathrm{tot}$ may be nonconvex in presence of heterogeneities, truncated CG detects negative curvature and return a step within the trust-region domain.

Moreover, $\Pi_\mathrm{tot} = \Pi_\mathrm{pot} + \Pi_\mathrm{dis}$ couples nonlocal elastic interactions along the crack front, through the $\fract$ and $\Hilbert$ operators involved in  $\Pi_\mathrm{pot}$ (see Eq.~\eqref{eq:Pi_pot_perturbed}), with local contributions from a heterogeneous fracture energy field, through $\Pi_\mathrm{dis}$ (see Eq.~\eqref{eq:Pi_dis}). When discretized in physical space, the corresponding Hessian is effectively dense because the nonlocal operators couple all front points. Moreover, material heterogeneity and mixed-mode effects introduce $\theta$-dependent coefficients and nonlinear products that break the translation-invariant structure of the homogeneous linear operator, leading to a coupling of the Fourier modes.

The minimization of $\Pi_\mathrm{tot}$ involves a large number of front degrees of freedom $a_i$ (typically thousands). To keep our method scalable, we implement the inner solver in a matrix-free fashion. Rather than assembling or storing the Hessian, the CG iterations only require Hessian-vector products. They can be evaluated efficiently via FFT, requiring $\mathcal{O}(N \log N)$ cost and $\mathcal{O}(N)$ memory per inner Krylov iteration, avoiding the $\mathcal{O}(N^3)$ factorization costs and $\mathcal{O}(N^2)$ memory associated with standard matrix methods.

Although our matrix-free approach removes the memory bottleneck, the number of Krylov iterations can still grow if the Hessian becomes poorly conditioned. To address this issue and seek further speed-up of our algorithm, we precondition the Newton-CG iterations in a matrix-free fashion, based on the inverse Hessian of an equivalent homogeneous problem.  The construction and derivation of this preconditioner are detailed in \ref{apdx:preconditioning}. This results in a speed-up of a factor $4-5$ for crack growth at low heterogeneity contrast and small $\nu$ (for shear cracks only). The gain decreases to a factor $2$ at maximum heterogeneity contrast, where strong spatial fluctuations of $\Gc$ along the diagonal of the Hessian matrix reduce the quality of the homogeneous approximation. It also decreases for larger Poisson’s ratio, where mixed-mode effects induce a non-diagonal coupling between Fourier modes. A natural extension of the present approach would be to design improved preconditioners that account for these effects, for example by constructing asymptotic inverses for heterogeneous systems with $\nu\neq 0$.

\subsection{Implementation in Python}

To satisfy the physical constraints described above while maintaining computational efficiency, we employ a (i) box-constrained Newton conjugate gradient method with a (ii) trust-region and a (iii) preconditioned matrix-free Krylov solver. The algorithm is implemented in Python using \texttt{PETSc} via the \texttt{petsc4py} interface \citep{petsc4py}, using the Bounded Newton Trust Region (\texttt{BNTR}) method. By dynamically tuning the \texttt{BNTR} gradient tolerance, we enforce a fixed tolerance of Griffith's criterion of at least $10^{-6}$ the value of the average fracture energy in the medium, tightened for weak disorder down to $10^{-3}$ of the heterogeneity contrast.

On top of the aforementioned requirements, we use dealiasing techniques to avoid spectral folding when evaluating the $0^\mathrm{th}$ Fourier coefficient of  double / triple product involved in Eqs.~\eqref{eq:Pi_pot_perturbed} and \eqref{eq:Pi_dis}. This makes the derivation of the gradient and Hessian product of the total energy less straightforward (but still manageable). For this reason, we leverage \texttt{JAX} \citep{jax} for automatic differentiation, to compute the gradient and Hessian-product of $\Pi_\mathrm{tot}$, and seek further speed-up with extended linear algebra and just-in-time compilation. Performance comparisons between the manually-derived, non-compiled Hessian--vector product implementation and its \texttt{JAX} counterpart show similar runtimes.

\subsection{Validation}
\label{subsec:validation}

In this section, we validate our numerical framework against newly derived first-order asymptotic analytical solutions for ``simple'' shear-rupture scenarios. These solutions describe stable crack propagation along a fracture plane with homogeneous or weakly heterogeneous fracture energy, and assume a small but nonzero Poisson’s ratio. Their derivation can be found in \ref{apdx:axisymetrical_solution}. Validation of the mode I formulation has also been conducted, but is omitted for brevity.

\subsubsection{Preliminaries}

Eq. \eqref{eq:Pi_pot_perturbed} can incorporate any loading conditions as long as the SIFs $\KIcirc$ and $\KIIcirc$, $\KIIIcirc$ for the circular crack configuration have a closed-form analytical solution or can be computed numerically. However, we have decided to restrict our implementation to situations where the stress acting on the rupture plane, in the absence of crack, only depends on the radial coordinate $r$.  In mode I, this corresponds to axisymmetric forces about the $(Oy)$ axis. In mixed modes {II}+{III}, it corresponds to loadings relevant e.g. to fluid injection along a planar frictional fault under constant shear \citep{saez_three-dimensional_2022}. The general SIF expressions for this class of axisymmetric loadings are given by Eq.~\eqref{eq:axisymmetric_SIF_1} and Eq.~\eqref{eq:axisymmetric_SIF_2+3}.

In this work, our simulations consider only two types of loading:
\begin{enumerate}
    \item a crack loaded by a pair of normal forces of intensity $P$ and opposite direction $\pm \textbf{e}_y$, applied at the center $O$ of an initial circular crack. The reference SIFs $\bar{K_p}$ for a penny-shaped crack of radius $a$ are given by \cite{tada_stress_2000} Eq.~(24.2):
    \begin{equation}
    \label{eq:SIFs_pointforce_P}
    \KIcirc(a,\theta) = \dfrac{P}{(\pi a)^{3/2}} \text{, and } \KIIcirc(a,\theta) = \KIIIcirc(a,\theta) = 0,
\end{equation}
    \item a crack loaded by a pair of shear forces of intensity $Q$ and opposite direction $\pm \textbf{e}_z$, applied at the center $O$ of an initial circular crack. The reference SIFs $\bar{K_p}$ for a penny-shaped crack of radius $a$ are given by \cite{tada_stress_2000} Eq.~(24.24):
    \begin{equation}
    \label{eq:SIFs_pointforce_Q}
        \KIcirc(a,\theta) = 0, \KIIcirc(a,\theta) = \dfrac{Q}{(\pi a)^{3/2}}\dfrac{1+\nu}{1-\nu/2}\cos{\theta} \text{, and } \KIIIcirc(a,\theta) = -\dfrac{Q}{(\pi a)^{3/2}}\dfrac{1-2\nu}{1-\nu/2}\sin{\theta}.
    \end{equation}
\end{enumerate}
All the validation results that will be presented thereafter in this section are of type 2. However, we report simulations of both type 1 \& 2 in Section~\ref{sec:applications}.

\subsubsection{Homogeneous material}

We first consider a homogeneous fracture energy field with $\Gc(r, \theta) = \Gc^0$. In mode I, the SIF $\KIcirc$, and thus the energy release rate $G$, are uniform along a circular front. An initially penny-shaped crack therefore remains circular as $P$ increases, as shown in Section~\ref{subsec:constraints}. Under shear loading, however, $\KIIcirc$ and $\KIIIcirc$, and hence $G$, vary with the polar angle $\theta$ (for any $\nu\neq 0$), so that even in a homogeneous medium, the front departs from circularity. This symmetry breaking has been reported in theory \citep{gao_nearly_1988}, in simulations of coplanar shear fractures \citep{favier_coplanar_2006}, and more recently for fluid-driven frictional shear ruptures \citep{saez_three-dimensional_2022}. Consequently, in the presence of shear stresses, finite- or boundary-element simulations cannot be reduced to an axisymmetric problem and are necessarily fully 3D. Perturbation-based approaches offer an alternative, as they allow one to numerically simulate the evolution of a non-circular crack front at reduced cost. They also offer sufficient analytical tractability to derive asymptotic solutions using the Poisson ratio $\nu$ as an expansion parameter.

Let $a(\theta)=a_0+\nu\,a_1(\theta)+\mathcal{O}(\nu^2)$, where $a_0$ is the radius of a penny-shaped crack propagating in a homogeneous brittle medium at $\nu=0$. Solving Griffith’s condition $G([a];\theta)=\Gc^0$ order-by-order, one obtains for point shear-force loading:
\begin{equation}
\label{eq:analytical_solution_homogeneous}
a(\theta, \nu) = a_0 \left(1 + \frac{\nu}{6} + \frac{\nu}{2}\cos(2\theta)\right) + \mathcal{O}(\nu^2) \text{, where } a_0 = \dfrac{1}{\pi}\left(\dfrac{Q^2}{E G_\mathrm{c}^0}\right)^{1/3}.
\end{equation}
The details of the derivation of Eq.~\eqref{eq:analytical_solution_homogeneous} are given in \ref{apdx:axisymetrical_solution}.

\begin{figure}[h]
	\centering
	\noindent\includegraphics[width=0.95\textwidth]{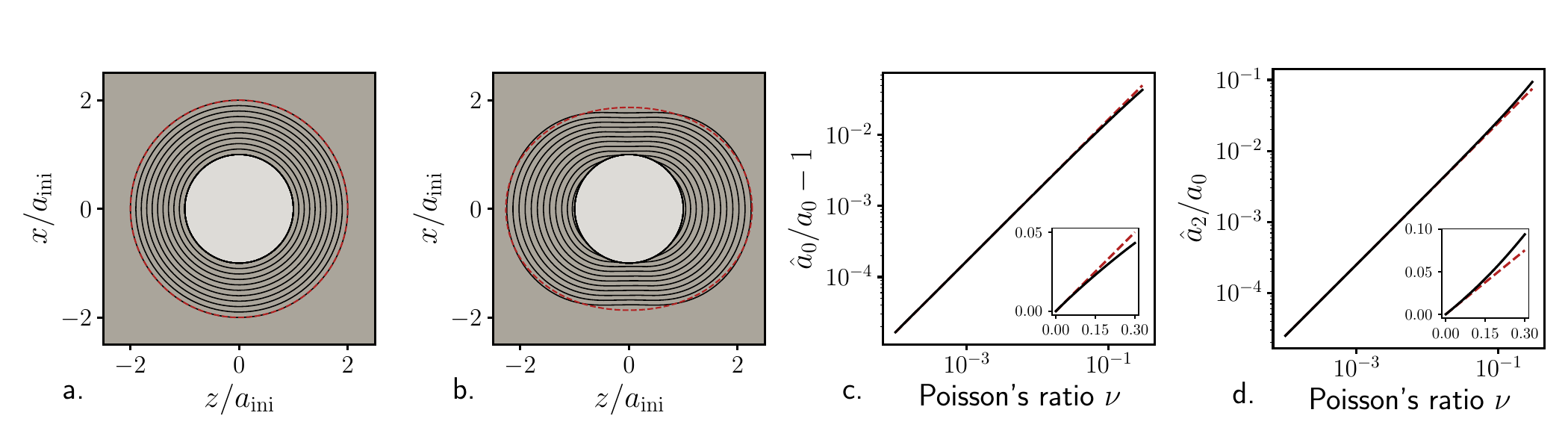}
	\caption{a.-b. A shear crack of initial radius $a_\mathrm{ini}$ (in solid bold black line) propagates within its plane $(xOz)$ under a pair of symmetric shear forces $Q$ applied at the center $O$. Consecutive crack front positions (in solid black lines) for Poisson's ratio $\nu = 0$. (a.) and $\nu = 0.2$. (b.) represent crack propagation in homogeneous fracture energy field as a result of increasing $Q$. The first order asymptotic solution  of Eq.~\eqref{eq:analytical_solution_homogeneous} (in red line) is compared to a crack front position at maximum computed $Q$. c. Relative difference of an average radius $\hat{a}_0$  with reference radius $a_0$ and d. Fourier coefficient $\hat{a}_2$ of the crack front position $a(\theta)$ at maximum computed $Q$. Numerical results (in solid black line) are compared to the first order asymptotic solution (in red dashed line).}
	\label{fig:validation_homogeneous}
\end{figure}

Figure~\ref{fig:validation_homogeneous} compares numerical results from our variational perturbative framework with the analytical prediction of Eq.~\eqref{eq:analytical_solution_homogeneous}. Solid black lines show successive front positions from an initial radius $a_\mathrm{ini}$ under increasing shear point force $Q$ for $\nu=0$ (Fig.~\ref{fig:validation_homogeneous}a) and $\nu=0.2$ (Fig.~\ref{fig:validation_homogeneous}b). The dashed red line is the analytical prediction at the final load. For $\nu=0$, the solution is axisymmetric, as expected. For $\nu=0.2$, the crack first advances near the lateral regions ($\theta=0^\circ,180^\circ$) while the remainder of the front remains pinned where $G([a];\theta)<\Gc^0$. At larger $Q$, the front approaches a quasi-elliptic, self-similar shape, owing to the lack of length scale in the problem. 

In Fig.~\ref{fig:validation_homogeneous}c and Fig.~\ref{fig:validation_homogeneous}d, we compare our numerical results with the analytical first-order solution of Eq.~\eqref{eq:analytical_solution_homogeneous} for the zero $\hat{a}_0$ and second $\hat{a}_2$ Fourier modes of $a(\theta)$. The gap between numerical and analytical results indicates when nonlinear terms matter. We see that our implementation follows the analytical linear trend in $\nu$ for both Fourier modes. We also observe that higher-order contributions start to become significant after $\nu = 0.1$, suggesting that higher-order theories are required to model crack growth in homogeneous media with practical values of Poisson's ratio $\nu \simeq  0.25-0.35$.

\subsubsection{Heterogeneous material}

Next, we move to the case where the shear crack propagates under point force loading $Q$ in a heterogeneous medium where $\Gc(r,\theta)$ fluctuates as:
\begin{equation}
    \Gc(r,\theta) = \Gc^0 \left(1 + \Delta \cos(p\theta)\right),
\end{equation}
where $\Gc^0$ is the average fracture energy, the real $\Delta$ is the heterogeneity amplitude and the positive integer $p$ is its mode. 

In that case, at first order, the contribution of the spatial variations of $\Gc$ superpose with the homogeneous solution of Eq. ~\eqref{eq:analytical_solution_homogeneous}, and the front position reads:
\begin{equation}
\label{eq:analytical_solution_heterogeneous}
\begin{aligned}
& a(\theta ) = a_0 \left(1 + \frac{\nu}{6} + \frac{\nu}{2}\cos(2\theta) - \frac{1}{p+3} \Delta \cos p \theta\right) + \mathcal{O}(\Delta^2, \nu^2, \Delta \nu) & \text { if } p>1, \\
\text{ and } & a(\theta) = a_0 \left(1 + \frac{\nu}{6} + \frac{\nu}{2}\cos(2\theta) - \frac{6}{5} \Delta \cos \theta\right) + \mathcal{O}(\Delta^2, \nu^2, \Delta \nu) & \text { if } p=1,
\end{aligned}
\end{equation}
where $a_0$ is the same as in Eq.~\eqref{eq:analytical_solution_homogeneous}. Details of the derivation for both mode I and mode {II}+{III} loading are given in \ref{apdx:axisymetrical_solution}.

\begin{figure}[H]
	\centering
	\noindent\includegraphics[width=0.9\textwidth]{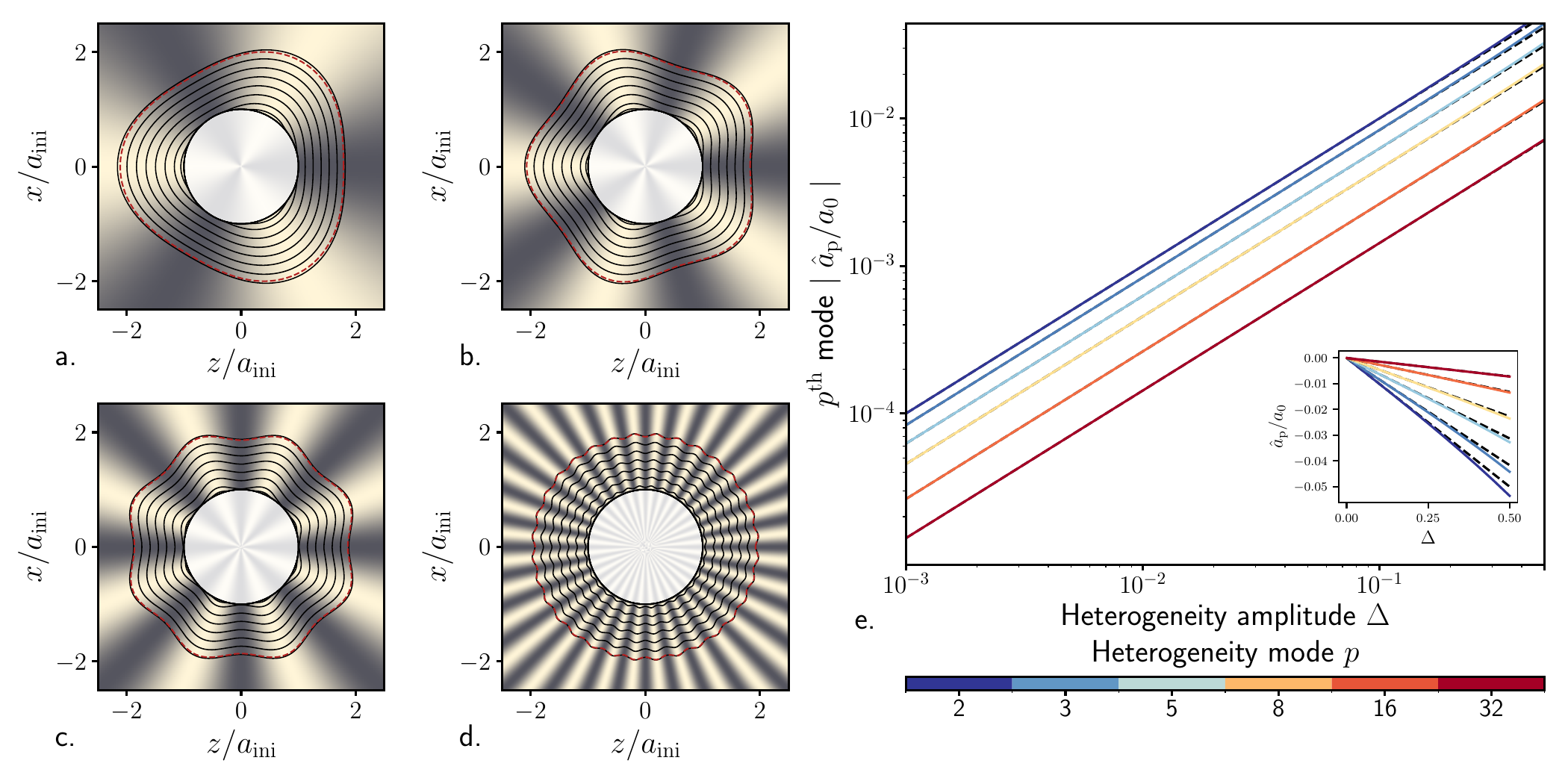}
	\caption{a.-d. A shear crack that follows the configuration of Figure~\ref{fig:validation_homogeneous} propagates in a heterogeneous fracture energy field $\Gc(\theta)$ (from beige to dark grey). Consecutive crack front positions (in solid black lines) for heterogeneity mode a. $p=3$; b. $p=5$; c. $p=8$ and d. $p=32$ represent crack propagation as a result of increasing point shear force $Q$. The first order asymptotic solution of Eq.~\eqref{eq:analytical_solution_heterogeneous} (in red line) is compared to a crack front position at maximum computed $Q$. e. Numerical results (in solid colored lines) and first-order asymptotic solution (in dashed black lines) for the $p^\mathrm{th}$ Fourier coefficient $\hat{a}_p$ of the crack front position $a(\theta)$ at maximum computed $Q$.}
	\label{fig:analytical_solution_heterogeneous}
\end{figure}

Figure~\ref{fig:analytical_solution_heterogeneous}a-d (black lines) shows crack-front evolution under increasing shear load $Q$ for various perturbation modes $p$, for $\nu = 0$ and $\Delta = 0.5$. As expected, crack advance starts in low-$\Gc$ regions. Under increasing $Q$, the crack front propagates as a whole and attains a steady-state self-similar regime, as in Section~\ref{subsec:validation}. Front deformations are dominated by the wavelength of the imposed $\Gc$ fluctuations and are most pronounced when the fluctuation wavelength is comparable to the crack radius (small $p$). Their amplitude decays approximately as $1/p$, consistent with Eq.~\eqref{eq:analytical_solution_heterogeneous}.

Looking at Fig.~\ref{fig:analytical_solution_heterogeneous}e, we see that our numerical result matches the theoretical Griffith-based first-order prediction, for the evolution of $p^{th}$ Fourier coefficient $\hat{a}_p$. Departures arise at larger heterogeneity contrasts $\Delta$ when $p$ is small, as front deformations are larger. For $\nu\neq 0$, the validity range of the first-order theory narrows further, as mode mixity introduces additional long-wavelength contributions. Together with the homogeneous case in the previous section, these comparisons validate both our variational perturbative framework and its numerical implementation.

\section{Illustrative applications}
\label{sec:applications}

In this section, we illustrate the potential of our method in tackling complex situations where the crack propagates along a fracture plane featuring arbitrary fluctuations of fracture energy. We give particular emphasis on the intermittency of crack growth and energy dissipation, as well as the apparent toughening arising from the presence of heterogeneities.

\subsection{Preliminaries}
\label{subsec: preliminaries_applications}

As above, loading consists of point-force pairs: normal forces of magnitude $P$ (mode~I, Eq.~\eqref{eq:SIFs_pointforce_P}) and shear forces of magnitude $Q$ (modes~{II}+{III}, Eq.~\eqref{eq:SIFs_pointforce_Q}). The fracture energy along the fracture plane is prescribed as an arbitrarily disordered field:
\begin{equation}
\label{eq: disordered_fracture_energy_field}
G_{\mathrm{c}}(r, \theta)=G_{\mathrm{c}}^{0} + \sigma f(r, \theta),
\end{equation}
where $G_{\mathrm{c}}^{0}$ is the average value of the fracture energy field, $\sigma$ its standard deviation and $f$ a fluctuation field with zero mean value and unit variance. The field is characterized by its probability density function $\mathbb{P}$, which captures the statistical distribution of local toughness, and its two-point correlation function $\mathcal{F}$, which describes its spatial distribution. Here we take $\mathbb{P}$ uniform on $[G_{\mathrm{c}}^{\min},G_{\mathrm{c}}^{\max}]$ and we consider isotropic exponential correlations with a single length scale $d$:
\begin{equation}
    \mathcal{F}(\mathbf{x}, \mathbf{x'}) = \mathrm{E}[f(\mathbf{x})f(\mathbf{x'})]= e^{-\rVert \mathbf{x} - \mathbf{x'} \lVert/d}
\end{equation}
where $\mathrm{E}[X]$ is the mathematical expectation of the random variable $X$, $\mathbf{x}=(r,\theta)$ and $\mathbf{x'}=(r',\theta')$ are two points along the fracture plane. An example of the fracture energy field is shown on the inset in Fig.~\ref{fig:disordered_example}a, together with its statistical descriptors $\mathbb{P}$ and $\mathcal{F}$ in Fig.~\ref{fig:disordered_example}d and Fig.~\ref{fig:disordered_example}e. More details on the field generation procedure are given in \citep{lebihain_size_2023}.

In the following, simulations are conducted for $29$ values of the disorder amplitude $\sigma / \Gc^0 \in(0,0.5]$, spanning nearly homogeneous to moderately heterogeneous fracture energy fields, with a maximum contrast $\Gc^{\max}/\Gc^{\min}\simeq 14$. Mode I simulations are rescaled for arbitrary Poisson ratio, whereas mode {II}+{III} simulations are restricted to $\nu\in{0,0.1,0.2}$, for which the first-order approximation remains valid. We generate $500$ fluctuation fields $f$. Each parameter pair $(\sigma/\Gc^0, \nu)$ is run for both $\Gc = \Gc^0 + \sigma f$ and $\Gc = \Gc -\sigma f$ for variance reduction purposes, resulting in a total of $1000$ realizations per parameter pair and 116,000 simulations in total. The load increments in $P$ and $Q$ are chosen so that, in the homogeneous reference case, the crack radius increases by a prescribed amount $\Delta a$ at each step, from an initial radius $a_\mathrm{ini}$ up to $a_{\max}$. Unless stated otherwise, simulations use $\Delta a=0.1\,d$, $a_\mathrm{ini}=0.001\,d$, $a_{\max}=100\,d$, trust-region radius $\Delta_{\mathrm{tr}}=0.1\,d$, and $N=2048$ discretization points along the front. Convergence with respect to $\Delta a$, $\Delta_{\mathrm{tr}}$, and $N$ is reported in \ref{apdx:convergence_study}.

\subsection{First example: Crack propagation in disordered medium}
\label{subsec:first_example}

In this first example, we highlight generic features of crack growth in heterogeneous media, with a particular focus on mixed-mode {II}+{III}, which has received less attention than mode~I in the literature. 

\begin{figure}[H]
	\centering
	\noindent\includegraphics[width=0.9\textwidth]{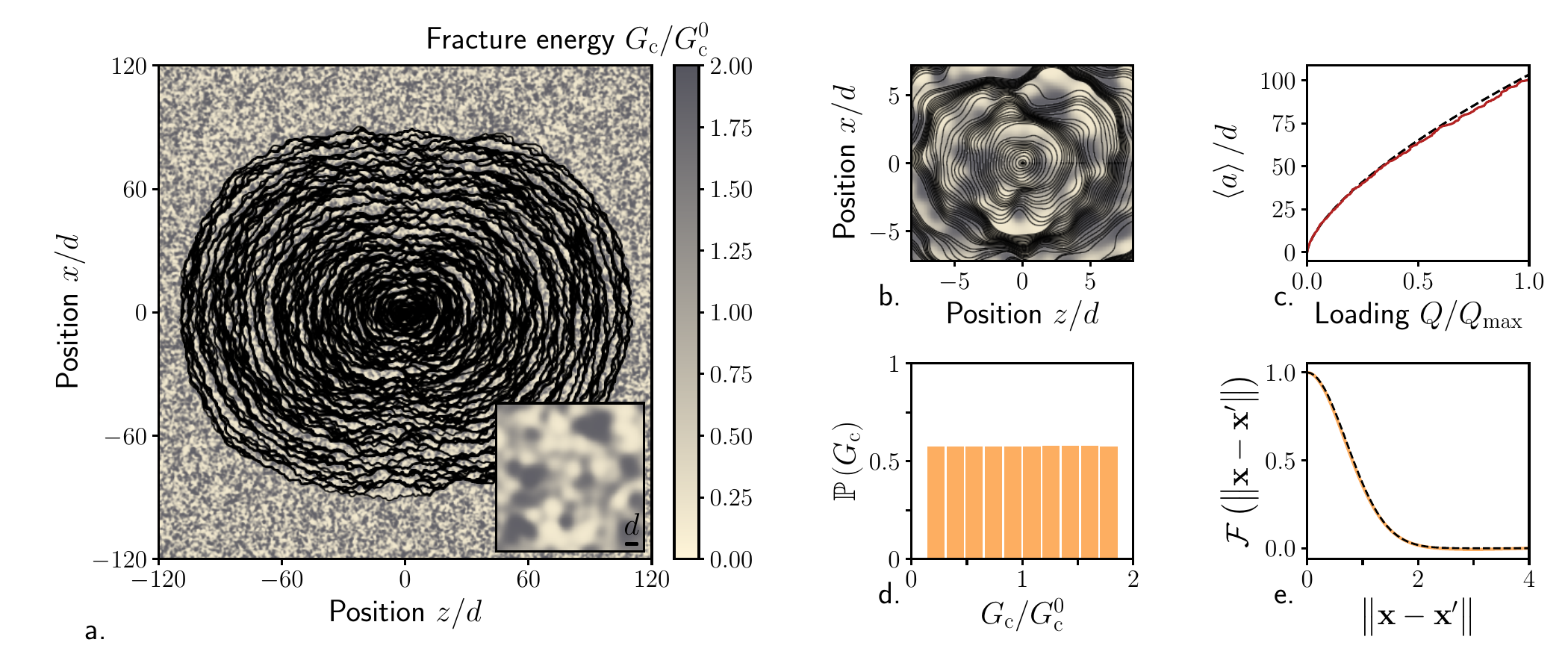}
	\caption{a. A shear crack propagates within its plane $(xOz)$ under a pair of symmetric shear forces $Q$ applied at the center $O$ in material with non-zero Poisson's ratio $\nu =0.2$. The disordered fracture energy field $\Gc(r,\theta)$ varies within the rupture plane over a characteristic length scale $d$ as shown on the inset in lower right corner. Successive equilibrium positions of the crack front during propagation are plotted in solid black lines. b. Early-time response that marks the transition between stable to intermittent crack growth, when the crack perimeter of approximate size of the Larkin length $\mathcal{P} \sim L_\mathrm{larkin}$ c. Normalized average crack radius $\langle a\rangle/d$ (solid red line), fluctuating around the reference radius (in dashed black line) of a homogeneous crack with $\Gc =\Gc^0$. d. Probability distribution function $\mathbb{P}$ of the local fracture energy. e. Two-point correlation function of the fracture energy field (in solid orange line) at typical heterogeneity scale $d$ that follows a prescribed bell-shaped distribution (in dashed black line).}
	\label{fig:disordered_example}
\end{figure}

Fig.~\ref{fig:disordered_example}a shows the propagation of the shear crack in a moderately heterogeneous fracture energy field with $\sigma/\Gc^0 = 0.5$, corresponding to $\Gc^{\max}/\Gc^{\min}\simeq 14$. The front positions are plotted in solid black lines every two loading increments for clarity. The dynamics of rupture growth qualitatively change with crack size. For small average crack radii ($\langle a \rangle \simeq d$; see Fig.~\ref{fig:disordered_example}b), crack propagation is continuous in time: the front advances everywhere as a whole, but exhibits large deformations because the crack encounters only few heterogeneities and long-wavelength perturbations dominate, consistent with Section 3.3.3. For larger cracks ($\langle a\rangle \simeq 5-100\,d$), growth becomes intermittent, alternating between pinning phases of quasi-static creep and avalanches during which portions of the front ''jump'' from one equilibrium configuration to another, as noticeable on Fig.~\ref{fig:disordered_example}a-c. This has been reported in experiments of coplanar crack propagation between sintered PMMA plates \citep{maloy_dynamical_2001, tallakstad_local_2011, lengline_interplay_2012} and patterned glass-PDMS interfaces \citep{chopin_morphology_2015}, as well as in numerical simulations \citep{bonamy_crackling_2008, laurson_avalanches_2010, patinet_quantitative_2013, bares_seismiclike_2019}. 

This transition is controlled by the ratio between the crack perimeter $\mathcal{P}$ and the Larkin length $L_\mathrm{larkin}\propto (\sigma/\Gc^0)^{-2} d$ \citep{tanguy_weak_2004}, which depends on the disorder intensity $\sigma/\Gc^0$ and the characteristic length $d$ of the fracture energy variations. When $\mathcal{P}\lesssim L_\mathrm{larkin}$ (small cracks and/or weak disorder), the front propagates in \textit{weak pinning} regime and advances smoothly as a whole. When $\mathcal{P}\gtrsim L_\mathrm{larkin}$ (large cracks and/or strong disorder), multiple metastable equilibria emerge and propagation becomes intermittent in the \textit{strong pinning} regime, as shown in Section 3.1.2. Note that traditional simulation approaches based on viscous regularization of Griffith’s criterion struggle to discriminate between weak and strong pinning regimes, which introduces bias in the emerging scale-free statistics \citep{laurson_avalanches_2010}. In contrast, our variational framework enables a direct assessment of crack stability, for instance via numerical stability/bifurcation analysis \citep{nguyen_stability_2000, leon_baldelli_numerical_2021}. Alternatively, one may perform a backward continuation at each step, reducing the load while relaxing irreversibility, to identify the local ``wells'' of stable equilibria.

\begin{figure}[H]
	\centering
	\noindent\includegraphics[width=0.9\textwidth]{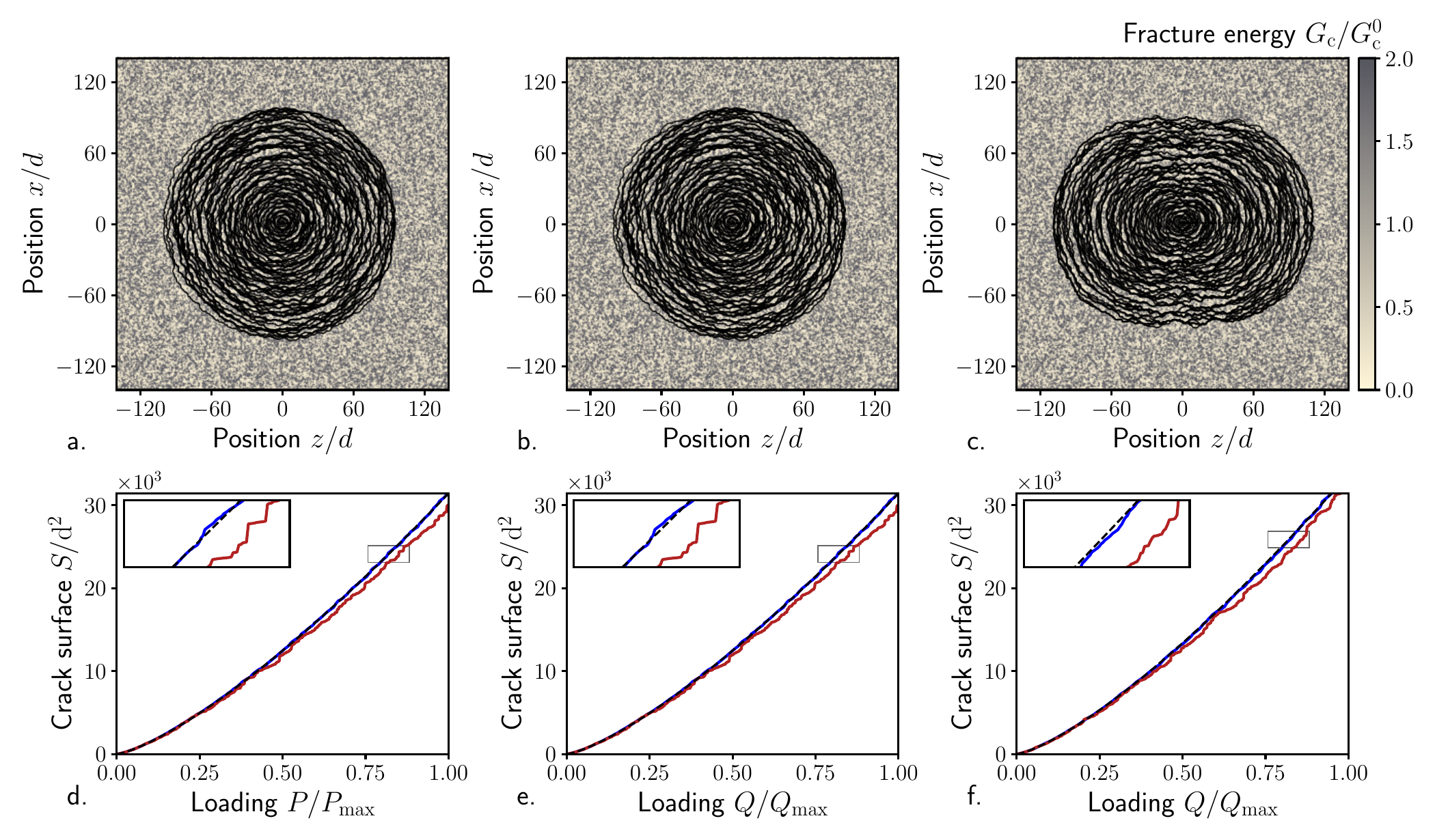}
	\caption{Crack growth in a heterogeneous material under different loading modes: a. mode I, b. mode {II}+{III} with $\nu = 0$, and c. mode {II}+{III} with $\nu = 0.2$ in a moderately disordered fracture energy field ($\sigma/\Gc^0 = 0.5$). d-f. The corresponding normalized crack surfaces  $S/d^2$ are plotted in dashed black line for a homogeneous field ($\sigma/\Gc^0 = 0$), in blue solid line for a weakly disordered field ($\sigma/\Gc^0 = 0.1$) and in red solid line for a moderately disordered field ($\sigma/\Gc^0 = 0.5$).}
	\label{fig:disordered_poisson}
\end{figure}

We investigate the role of mode mixity in Fig.~\ref{fig:disordered_poisson} by comparing crack propagation on the same toughness field in three configurations: mode~I (Fig.~\ref{fig:disordered_poisson}a), mixed mode {II}+{III} with $\nu=0$ (Fig.~\ref{fig:disordered_poisson}b), and mixed mode {II}+{III} with $\nu=0.2$ (Fig.~\ref{fig:disordered_poisson}c). We also report the evolution of fractured surface $S= \int_0^{2\pi} a(\theta)^2/2\, \mathrm{d}\theta$ with increasing loading for the three cases (Fig.~\ref{fig:disordered_poisson}d–f). Mode~I and mixed mode {II}+{III} with $\nu=0$ yield identical results. Indeed, in the limit $\nu=0$ and for equal force amplitudes $P=Q$, the energy release rate $G$ in mode {II}+{III} reduces to its mode I expression, so the same $\Gc$ field produces the exact same propagation history.

For $\nu\neq 0$, the front develops a rough quasi-elliptic shape, which is consistent with Section~3.3.2. This behavior is also consistent with the analysis of \cite{gao_penetration_1991}, which shows that the crack-front stiffness differs between modes, with mode~{III} being stiffer than mode~{II}, and mode~I being intermediate. In our geometry, the upper and lower sectors are predominantly loaded in mode~III and therefore advance more slowly than the lateral sectors loaded in principal mode~II, yielding a quasi-elliptical crack front. We do not observe a clear shift in the onset of intermittency in the fracture surface evolution (Fig.~\ref{fig:disordered_poisson}d–f), although a rigorous assessment would require a dedicated nonlinear stability analysis at each computed equilibrium position.
    
\subsection{Second example: Intermittent energy dissipation}
\label{subsec:second_example}

In the second example, we focus on the intermittency of energy dissipation. Regardless of loading mode, radial fluctuations of fracture-energy induce unstable crack propagation, characterized by slip/opening bursts spanning several orders of magnitude. Such intermittent dynamics are generic in rupture-like processes \citep{bonamy_failure_2011, laurson_evolution_2013}, including adhesive crack propagation in soft solids \citep{sanner_crack-front_2022}, frictional slip and associated seismicity on heterogeneous faults \citep{ben-zion_collective_2008,candela_fault_2011, dublanchet_interactions_2013}, and brittle fracture of disordered materials \citep{schmittbuhl_interfacial_1995,rosso_roughness_2002, patinet_quantitative_2013}. These phenomena demonstrate that microscale disorder substantially changes the macroscopic response relative to a homogenized description, motivating models that are both numerically efficient and capable of capturing intermittency.

\begin{figure}[H]
		\centering
		\noindent\includegraphics[width=0.9\textwidth]{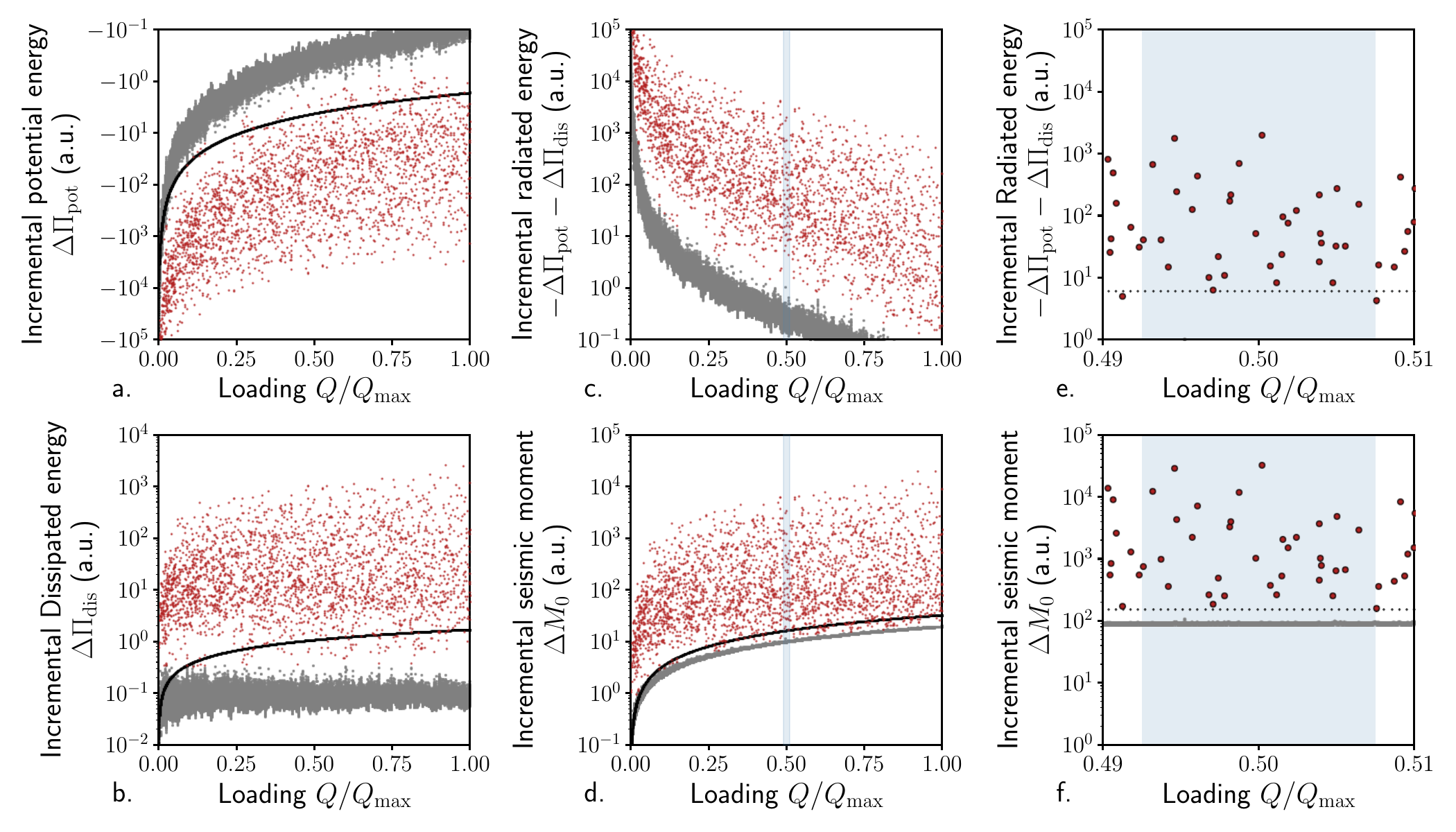}
		\caption{ Fluctuations of a. an incremental potential energy $\Delta \Pi_\mathrm{pot}$; b. incremental dissipated energy $\Delta \Pi_\mathrm{dis}$; c. the difference between the incremental change in potential energy $\Delta \Pi_\mathrm{pot}$ and incremental dissipated energy $\Delta \Pi_\mathrm{dis}$ with opposite sign, that represents the amount of energy radiated in the bulk per instability event and d. the zero-order incremental seismic moment $M_0$ in a disordered fracture energy field with $\sigma/\Gc^0 = 0.5$. e. and f. are zoom on panels c. and d. around $Q/Q_\mathrm{max} \simeq 0.5$. All quantities are presented in arbitrary units and compared with the reference ones obtained for a homogeneous field (in black dots). Stable propagation events (in grey dot markers) are distinguished from the unstable ones (in red dot markers).}
		\label{fig:intermittency}
	\end{figure}

We illustrate this behavior using the simulation of the previous example ($\sigma/\Gc^0=0.5$) and the homogeneous case. Then we compute the changes in \textit{potential} and \textit{dissipated} energies from Eqs.~\eqref{eq:Pi_pot_perturbed} and \eqref{eq:Pi_dis}, as the shear force $Q$ increases. Figure~\ref{fig:intermittency}a shows the increment $\Delta\Pi_\mathrm{pot}$ between two loading steps $Q_{n}$ and $Q_{n+1}$. Here, $\Delta\Pi_\mathrm{pot}$ is negative because our method computes changes in $\Pi_\mathrm{pot}$ with crack advance, excluding the baseline potential energy $\Pi_\mathrm{pot}^\mathrm{uncracked}$ of the uncracked configuration, which increases with loading. During pinning phases (in grey dot markers), the system stores elastic energy, and $\Pi_\mathrm{pot}$ is above the homogeneous trend (in solid black line). When the crack unpins (in red dot markers), $\Pi_\mathrm{pot}$ drops suddenly, and stored energy is either radiated within the elastic body or dissipated by creating new free surfaces. 

Figure~\ref{fig:intermittency}b shows the change $\Delta\Pi_\mathrm{dis}$ in dissipated energy between two loading steps. While the crack is pinned and propagates in a subcritical regime (in grey dot markers), dissipation increases slowly and remains orders of magnitude below the homogeneous case. During instabilities (in red dot markers), it exhibits sharp spikes associated with the sudden increase in crack surface. However, these bursts remain generally smaller than the potential energy drops, indicating that some energy has been radiated into the bulk. The observed intermittency in both $\Pi_\mathrm{pot}$ and $\Pi_\mathrm{dis}$ is correlated and primarily controlled by fluctuations in the local fracture energy field, which cause episodic elastic energy storage and release. Importantly, both $\Delta\Pi_\mathrm{pot}$ and $\Delta\Pi_\mathrm{dis}$ exhibit multiscale fluctuations spanning several orders of magnitude, even though the fracture-energy field is bounded and correlated over a single length scale $d$. This variability arises from long-range elastic interactions along the front, which couple distant segments during pinning and depinning. The resulting crackling statistics are reminiscent of the scale-free event distributions observed in earthquake rupture \citep{scholz_mechanics_2019, bak_unified_2002, ben-zion_collective_2008}.

To complete the energy balance \citep{jestin_energy_2019, fialko_fracture_2015}, our simulation would require either to include dynamic inertial effects or to know the baseline value of the potential energy of an uncracked body $\Pi_\mathrm{pot}^\mathrm{uncracked}$. However, one may argue that during instabilities (in red dot markers), the change in total potential energy is dominated by the incremental change due to crack advance, since the loading increment remains relatively small. Hence, the incremental change in energy radiated in the bulk can be assessed through the sum of $-\Delta\Pi_\mathrm{pot}$ and $-\Delta\Pi_\mathrm{dis}$ (Fig.~\ref{fig:intermittency}c). This quantity has no relevant physical meaning during pinning phases (in grey dot markers). 

Earthquake source studies typically rely on the moment-rate function to describe the spatiotemporal history of slip on a fault \citep{lay_modern_1995}, since seismicity is characterized by the intermittent character of the radiated energy and consequently as sudden bursts in the moment rate function. In that spirit, our setup can be viewed as an idealized double-couple source, without explicitly representing the shear tractions arising from friction and inertia (see Section~3.2 in \cite{aki_quantitative_2009}). For the shear point-force loading of Eq.~\ref{eq:axisymmetric_point_force}, the associated (symmetric) moment tensor takes the form:
\begin{equation}
\label{eq:moment_tensor}
    \mathbf{M}=\left(\begin{array}{ccc}0 & 0 & M_0 \\ 0 & 0 & 0 \\ M_0 & 0 & 0\end{array}\right),
\end{equation}
where $M_0$ is the cumulative seismic moment:
\begin{equation}
\label{eq:moment_tensor_component}
    M_0(Q_n) =\mu \int_{0}^{2\pi}\int_{0}^{a(\theta)}  \Delta u_\mathrm{z}([a];r, \theta) \, r \mathrm{d}r \mathrm{d}\theta,
\end{equation}
and $\mu = \frac{E}{2(1+\nu)}$ is the shear modulus. 

Computing $M_0(Q_n)$ for an arbitrary front $a(\theta)$ requires the slip $\Delta u_\mathrm{z}([a];r,\theta)$ all over the perturbed crack surface. To stay consistent with the first-order perturbation theory of Section~\ref{sec:theory}, one would need to calculate the perturbation of crack shear displacement using  \cite{gao_nearly_1988}’s Eq.~(1), from its reference value for the penny-shaped crack, obtained by \cite{fabrikant_applications_1989} in Eq.~(4.4.52). However, for the sake of simplicity, we use a zero-order estimate based on the circular-crack solution, in which $M_0$ depends only on the average front radius $\hat{a}_0$:
\begin{equation}
\label{eq:moment_tensor_component_approx}
M_0(Q_n) \simeq \frac{4Q_n}{\pi}\,\frac{1-\nu}{2-\nu}\,\hat{a}_0 + \mathcal{O}(\norm{\Delta a}).
\end{equation}

In Fig.~\ref{fig:intermittency}d, we show the increment of cumulative seismic moment $ \Delta M_0$ between two loading steps. $\Delta M_0$ can be interpreted as a proxy for the moment-rate function. A key advantage of our variational energetic formulation is that we can access to a proxy for radiated energy (red dot markers on Fig.~\ref{fig:intermittency}c), through the changes of $\Pi_\mathrm{pot}$ and $\Pi_\mathrm{dis}$ during instabilities, and compared to the burst-like moment release (red dot markers in Fig.~\ref{fig:intermittency}d), which, in a dynamic setting, would be associated with energy being radiated as bulk waves during rapid frictional rupture \citep{kanamori_diversity_2004}.

To facilitate comparison, Fig.~\ref{fig:intermittency}e–f zooms into the same window around $Q/Q_\mathrm{max} \simeq 0.5$. Peaks in $\Delta M_0$ and in the radiated energy occur at the same time, indicating a clear event-by-event correspondence. However, $\Delta M_0$ tends to increase with crack size, whereas the energy-change measure does not. A likely explanation is that $\Pi_\mathrm{pot}$ is defined relative to the uncracked reference energy $\Pi_\mathrm{pot}^\mathrm{uncracked}$, which increases continuously with loading and can bias incremental energy changes, especially when, as in our case, the load increments vary to maintain a constant increase in the reference radius ($\Delta Q \propto \sqrt{a_0} \, \Delta a$).

We interpret Fig.~\ref{fig:intermittency}c and~\ref{fig:intermittency}d as follows: the isolated, high-magnitude moment or radiated energy releases (in red dot markers) correspond to episodes of large slip, consistent with unstable crack propagation and the generation of seismic events, whereas clusters of low-magnitude events (in grey dot markers) reflect limited slip increases that can be interpreted as aseismic, stable sliding. 

Such intermittent rupture events are widely reported in heterogeneous fracture and fault models \citep{ben-zion_earthquake_1993, schmittbuhl_interfacial_1995, maloy_local_2006,  bares_seismiclike_2019, almakari_fault_2026}. Here, the key point is that a bounded, single-scale fracture energy field is sufficient to generate complex sequences of moment release.

\subsection{Third example: material toughening}
\label{subsec:third_example}

As a final example, we discuss how microscopic fluctuations of fracture energy can change the apparent fracture resistance at the macroscopic scale. Toughening by crack pinning on periodic arrangements of tough inclusions was first addressed by \cite{gao_first-order_1989} using first-order perturbation theory and extended to all orders by \cite{bower_bridging_1991}, with successful comparison to experiments performed by \cite{mower_experimental_1995}. Later contributions have focused on disordered interfaces with random fluctuations of the fracture energy \citep{roux_effective_2003, patinet_quantitative_2013, demery_microstructural_2014}. For semi-infinite cracks, these studies showed systematic disorder-induced toughening, whose intensity scaled as $(\sigma/\Gc^0)^2$ from the baseline set by the average fracture energy $\Gc^0$ (simple mixture rule). A simple interpretation is provided in Fig.~\ref{fig:trust_region}b: due to instabilities, the crack does not sample the fracture energy field uniformly, but preferentially visits the ``peaks'' of the fracture energy distribution, causing an apparent toughening at the macroscale. However, in 3D, the actual fracture energy distribution is not that of the fracture energy field, but the distribution sampled by the pinned crack front, biased toward the largest $\Gc$ values \citep{patinet_quantitative_2013}.

Here, we discuss the impact of finite-size effects and mode mixity on the apparent fracture resistance. We do not attempt to define a unique effective fracture energy \citep{hossain_effective_2014, lebihain_effective_2021, michel_merits_2022}. Instead, we focus on the crack surface area variation $\Delta S$: 

\begin{equation}
\label{eq:area_variation}
    \Delta S = \dfrac{1}{2}\int_0^{2\pi}a(\theta)^2 \mathrm{d}\theta - S_0,
\end{equation}
which corresponds to the difference between the actual crack surface area and its reference value $S_0$ for a crack propagating in a homogeneous material of fracture energy $\Gc^0$. By construction, $\Delta S=0$ for $\Gc = \Gc^0$, no matter $\nu$. Intuitively, $\Delta S < 0$ at fixed load $P$ or $Q$ indicates toughening from the mixture rule, as the crack could not grow as much as it would in a homogeneous material of fracture energy $\Gc^0$. In contrast, $\Delta S >0$ indicates weakening.

Figures~\ref{fig:disordered_poisson}d-f show that $\Delta S$ is negative and grows (in absolute value) with increasing disorder intensity $\sigma/\Gc^0$, meaning that fluctuations of fracture energy increase fracture resistance, in accordance with results obtained for semi-infinite cracks \citep{roux_effective_2003, patinet_quantitative_2013, demery_microstructural_2014}. This toughening is mainly associated with the storage of elastic energy during the pinning phase (Fig.~\ref{fig:intermittency}a), in line with previous numerical \citep{hossain_effective_2014, xiong_pinning_2024} and experimental \citep{liu_high_2020, triclot_toughening_2024} observations.

\begin{figure}[h]
	\centering
	\noindent\includegraphics[width=0.9\textwidth]{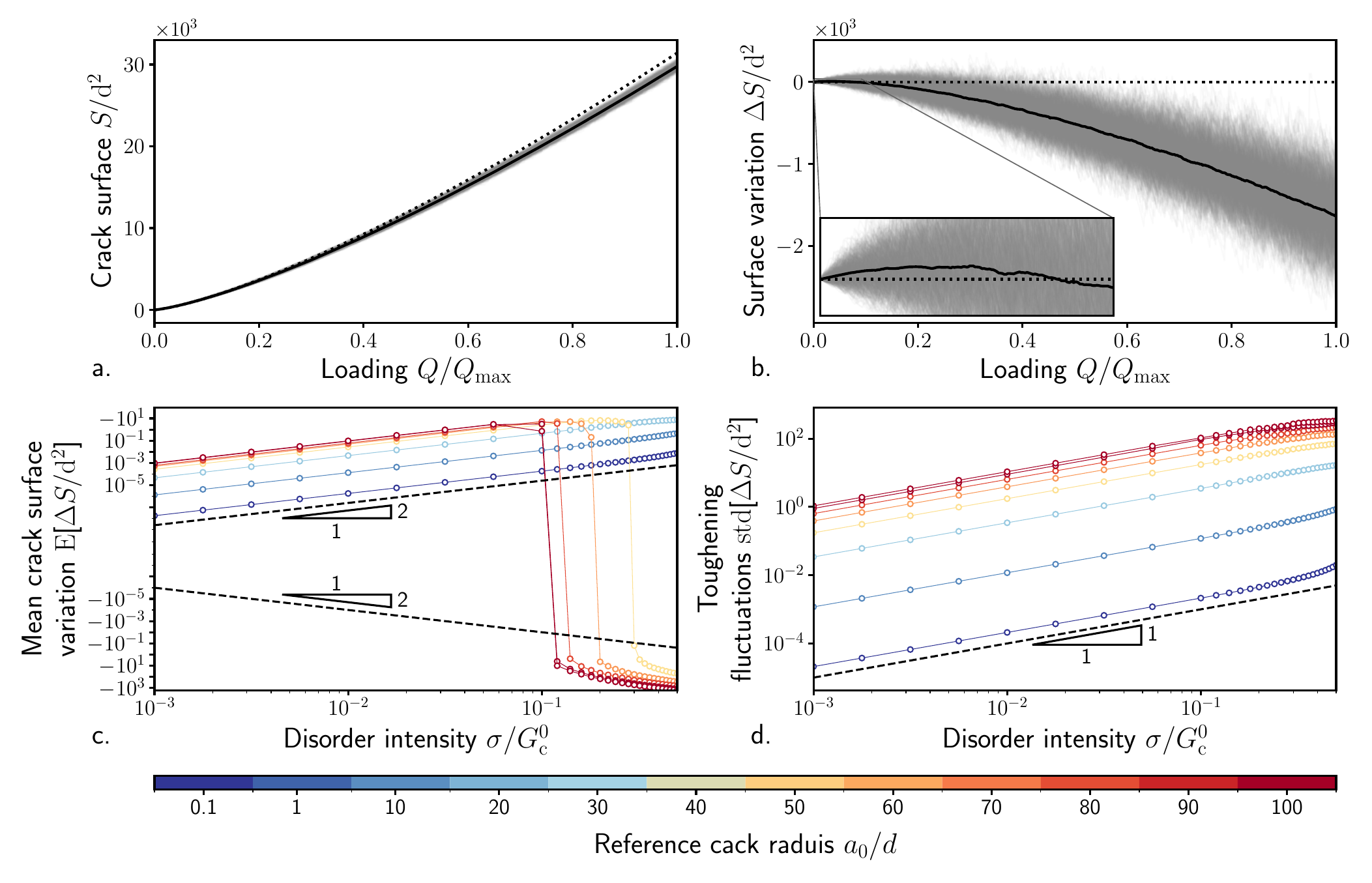}
	\caption{a. Normalized crack surface area $S/d^2$ and b. Surface variation $\Delta S/d^2$ for mixed mode II+III ($\nu=0,\, \sigma/\Gc^0=0.5$) crack propagation under increasing shear force $Q$ ($Q_\mathrm{max}$ corresponds to $a_0 \simeq 100\,d$). Thin grey solid lines represent individual realizations, the solid black line is the average over $1000$ realizations, and the dotted black line shows the homogeneous reference. The inset shows transition from toughening to weakening regime. c. Mathematical expectation $\mathrm{E}[\Delta S/d^2]$ and d. standard deviation $\mathrm{std}[\Delta S/d^2]$ of the surface variation as a function of the disorder intensity $\sigma/\Gc^0$ for several values of loading $Q$, represented as reference radius $a_0$ (from blue to red colored lines).}
	\label{fig:homogenization_scalings}
\end{figure}

To quantify this toughening, we analyze the full dataset of 116,000 simulations. Recall that we sample $\sigma / \Gc^0 \in(0,0.5]$, and for each $\sigma / \Gc^0$, we generate $500$ independent field fluctuations $f$ and conduct simulations for the two antithetic fields $\Gc = \Gc^0 + \sigma f$ and $\Gc = \Gc -\sigma f$, for variance reduction, yielding $1000$ realizations per parameter set. The evolution of crack surface area $S$ with increasing shear force $Q$ (mixed mode {II}+{III}, $\nu = 0$ and $\sigma/\Gc^{0}=0.5$) is shown in Fig.~\ref{fig:homogenization_scalings}a, together with the corresponding $\Delta S$ in Fig.~\ref{fig:homogenization_scalings}b. Thin grey lines show individual realizations, the solid black line is the ensemble mean over $1000$ realizations, and the dotted black line is the homogeneous reference ($\Gc=\Gc^{0}$). While individual realizations can exhibit either apparent weakening or toughening, mean trends are noticeable depending on the crack size. For small cracks ($a_0 \simeq 5\,d$, corresponding to $Q/Q_\mathrm{max} \leq 0.01$, see insets of Fig.~\ref{fig:homogenization_scalings}a and~\ref{fig:homogenization_scalings}b), the presence of heterogeneities results in an overall weakening of the material ($\Delta S \lesssim  0$), while large cracks ($a_0 \simeq 50\,d$, $Q/Q_\mathrm{max} \geq 0.4$) show unconditional toughening ($\Delta S < 0$). 

The small-$a$ regime reflects the tendency of cracks to bow more in zones of low fracture energy than in the tough regions, thus lowering the fracture energy sampled by the front \citep{vasoya_geometrically_2013, lebihain_size_2023}. Interestingly, this effect appears in higher-order theories, but our numerical method, which is first-order in $G$ but effectively ``all-order'' in $a$, captures this effect qualitatively. The large-$a$ regime is controlled by the onset of instabilities in the strong pinning regime \citep{roux_effective_2003}, and is also of second-order in $\sigma/\Gc^0$, as shown in previous works \citep{patinet_quantitative_2013, demery_microstructural_2014}.

To assess how $\Delta S$ scales with $\sigma/\Gc^0$, we estimate its mathematical expectation $\mathrm{E}[\Delta S]$, and standard deviation $\mathrm{std}[\Delta S] = \sqrt{\mathrm{E}[\Delta S^2] - \mathrm{E}[\Delta S]^2}$ using ensemble averages. \cite{lebihain_size_2023} showed, using a second-order theory, that $\mathrm{E}[\Delta S] \propto (\sigma/\Gc^0)^2$ while $\mathrm{std}[\Delta S] \propto \sigma/\Gc^0$. As a result, resolving the quadratic mean $\mathrm{E}[\Delta S]$ requires a large number $n$ of realizations, as the sampling error on $\mathrm{E}[\Delta S]$ scales as $\mathrm{std}[\Delta S] / \sqrt{n} = (\sigma/\Gc^0)/ \sqrt{n}$. \cite{lebihain_size_2023} avoided this issue by using variance reduction methods and estimating $\mathrm{E}[\Delta S]$ from ensemble averages performed on antithetic pairs of fracture energy fields $\Gc = \Gc^0 + \sigma f$ and $\Gc = \Gc -\sigma f$, for each field $f$ (zero mean, unitary variance):
\begin{equation}
\label{eq: effective_toughness_variance_reduction}
\mathrm{E}[\Delta S] \simeq \frac{\sum_{f} \Delta S[+f]+\Delta S[-f]}{2 n} 
\end{equation}
The standard deviation $\mathrm{std}[\Delta S]$ is computed from the runs $\Gc = \Gc^0 + \sigma f$ only. Figure~\ref{fig:homogenization_scalings}c \& d. show $\mathrm{E}[\Delta S] \propto (\sigma/\Gc^0)^2$ and $\mathrm{std}[\Delta S] \propto \sigma/\Gc^0$, consistent with previous findings \citep{patinet_quantitative_2013, demery_microstructural_2014, lebihain_size_2023}. As discussed before, heterogeneities drive material weakening ($\mathrm{E}[\Delta S] > 0$) for small cracks and toughening ($\mathrm{E}[\Delta S] < 0$) for large cracks. This transition shifts to smaller cracks for larger disorder intensity $\sigma/\Gc^0$, consistent with a transition from weak to strong pinning when the crack perimeter $\mathcal{P}$ grows larger than the Larkin length $L_\mathrm{larkin} \propto (\sigma/\Gc^0)^{-2}d$. The results for mode~I are identical, as explained in Section 4.2, and those in mixed mode {II}+{III} with $\nu = 0.2$ showed no clear difference. 

Finally, because $\mathrm{E}[\Delta S]$ scales quadratically with disorder intensity $\sigma/\Gc^0$, predicting its prefactor quantitatively would require pushing our simulations to second order \citep{leblond_second-order_2012, lebihain_size_2023}.

\section{Conclusion}
\label{sec: Conclusion}

In this article, we developed a variational approach for mixed-mode coplanar propagation of \textit{sharp} cracks in heterogeneous brittle media, connecting the formulation of \cite{francfort_revisiting_1998} with \cite{rice_first-order_1985}'s perturbation theory. Our core contribution is the theoretical derivation of the asymptotic expansion of the potential energy of a quasi-circular crack in Section~\ref{sec:theory}, in relation with traditional expansion of the stress intensity factors and energy release rate \citep{gao_somewhat_1987, gao_nearly_1988}. Next, we proposed in Section~\ref{sec:numerics} a numerical implementation that combines Fast Fourier Transforms for rapid evaluation of the potential energy with a matrix-free box-constrained Newton conjugate gradient solver, with trust region and physics-based preconditioning, to compute equilibrium crack-front positions through nonconvex minimization. The algorithm enforces the key ingredients of brittle fracture in heterogeneous media (irreversibility, energy barriers, and long-range elastic interactions), while maintaining numerical efficiency. Our \texttt{Python} implementation, based on \texttt{petsc4py}, has been validated against new 3D analytical benchmarks, derived from the perturbation theory.

Finally, we demonstrate the potential of our method in Section~\ref{sec:applications}, by revisiting crack growth in disordered media, but accounting for finite-size effects and mode mixity. Namely, we were able to model crack propagation in disordered materials containing tens of thousands of heterogeneities, with a heterogeneity contrast up to $\Gc^\mathrm{max}/\Gc^\mathrm{min} = 14$, within hours in a single core computer. Our simulations showed a transition from stable to intermittent crack growth, consistent with previous findings obtained for semi-infinite cracks. Mode mixity has limited influence on the onset of intermittency, but it induces a quasi-elliptic front shape in mixed mode {II}+{III}. We further show how intermittency redistributes dissipation in time, with implications for rupture dynamics in geophysical settings. Finally, we quantify the impact of heterogeneity on apparent macroscopic resistance, highlighting a crack-size-dependent crossover from disorder-induced weakening to toughening relative to a mixture-rule baseline, controlled by the onset of instabilities. This mechanism suggests ways to design damage-tolerant materials through controlled R-curve behavior by microscale patterning.

Our method can be readily adapted to other crack geometries, including semi-infinite cracks \citep{rice_first-order_1985, gao_shear_1986}, circular contacts \citep{gao_nearly_connections_1987, gao_linear_1989}, tunnel cracks \citep{leblond_tensile_1996, lazarus_three-dimensional_1998}, and extended to higher-order theories, whenever a second-order asymptotic expansion of the energy release rate $G$ is accessible \citep{leblond_second-order_2012, lebihain_size_2023}. A natural next step is cohesive fracture, for which both perturbative descriptions \citep{lebihain_cohesive_2022, roch_dynamic_2023} and a variational structure \citep{dalmaso_cohesive_2007} exist, to include size-dependency in material failure and model the entire fracture process, from initiation to propagation. The framework can further be generalized to dynamic rupture \citep{morrissey_perturbative_2000, kolvin_comprehensive_2024} to study wave–front interactions beyond the stationary setting of \cite{kolvin_dual_2025}. Finally, similar ideas may be applied to other interfacial problems, including contact between a rigid indenter and a rough \citep{argatov_controlling_2021, sanner_why_2024} or chemically-patterned \citep{sanner_crack-front_2022, argatov_mechanics_2023} soft material, or frictional ruptures generated by the injection of fluids in the subsurface \citep{saez_three-dimensional_2022}.

\section*{Acknowledgements}
ML acknowledges the support of the French Agence Nationale de la Recherche (ANR), under grant ANR-24-CE08-5174 (project DURABLE). Additionally, the authors are grateful to Jeremy Bleyer and Pierre Dublanchet for their valuable feedback during the working process.

This research was funded, in whole or in part, by French Agence Nationale de la Recherche (ANR), under grant ANR-24-CE08-5174. A CC-BY public copyright licence has been applied by the authors to the present document and will be applied to all subsequent versions up to the Author Accepted Manuscript arising from this submission, in accordance with the grant’s open access conditions.\\

\section*{Data Availability}

The source codes used to produce all numerical results presented in this article are openly available at \url{https://github.com/SerafimEgorov/MMH-BriEF-Variational-Solver}. The repository contains Python scripts for: i) the propagation of tensile and shear cracks in a heterogeneous fracture energy field; ii) the generation of a disordered fracture energy field and its pre-computation. Raw data from $116,000$ individual simulations are not archived but can be fully regenerated from the provided scripts using the parameter sets reported in the article.

\noindent \includegraphics[width=3cm]{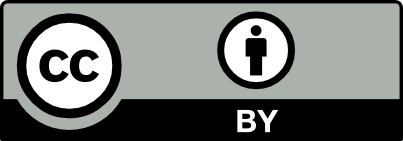} \href{https://creativecommons.org/licenses/by/4.0/}{\textbf{Distributed under a Creative Commons Attribution | 4.0 International licence}}

\section*{Declaration of Generative AI and AI--assisted technologies in the writing process}

    During the preparation of this work, the authors used ChatGPT (OpenAI) and Claude (Anthropic) to optimize the quality of the text. After using this tool, the authors reviewed and edited the content as needed and take full responsibility for the content of the publication.

\appendix

\section{Preconditioning strategy for the Newton conjugate gradient method}
\label{apdx:preconditioning}

To solve Eq.~\eqref{eq:variational_formulation}, we employ a box-constrained Newton conjugate gradient (CG) method \citep{nocedal_numerical_2006}. At each Newton iteration, we solve, on the working set (i.e., indices where the bound constraints are not active), the following system:
\begin{equation}
\label{eq:newton_cg}
H_{ij} v_j = - g_i,
\end{equation}
where $\mathbf{H}=(H_{ij})$ is the discrete Hessian of the total energy and $\mathbf{g}=(g_i)$ its gradient.

We solve Eq.~\eqref{eq:newton_cg} iteratively, which requires Hessian vector product $H_{ij}v_j$ only in a matrix-free formulation. The convergence of Newton CG depends on the spectral properties of $\mathbf{H}$, in particular the distribution of its eigenvalues \citep{nocedal_numerical_2006}. Considering the case of a mode~I or mode {II}+{III} ($\nu = 0$) penny-shaped crack of radius $a_0$ and energy release rate $\bar{G}$ propagating in a homogeneous medium of fracture energy $\Gc^0$, the Hessian vector product can be written as:
\begin{equation}
\label{eq:homogeneous_hessian}
    H_{ij}^\mathrm{hom} v_j=- \frac{2 \pi}{N} (\bar{G}(a_0)+a_0\bar{G}'(a_0)-\Gc^0) v_j+\frac{2 \pi}{N} G_0\left(\hat{a}_0\right) \fract[v]_j,
\end{equation}
where $\fract$ is the square-root Laplacian defined in Eq. \ref{eq:Operator_L}. Since $\fract$ is diagonal in Fourier space, $H^\mathrm{hom}$ is diagonal in Fourier space as well, with eigenvalues:
\begin{equation}
\label{eq:eigenvalues_hessian}
    \lambda_k = -\frac{2\pi}{N} \left((1-\abs{k})\bar{G}(a_0)+a_0\bar{G}'(a_0)-\Gc^0\right)
\end{equation}
Thus, in absence of heterogeneities, the eigenvalues of $\mathbf{H}=\mathbf{H}^\mathrm{hom}$ are uniformly distributed, which degrades the conditioning of the Newton system and slows CG convergence.

We therefore use a physics-based preconditioner $\mathbf{P}$ based on the inverse of the homogeneous Hessian $H^\mathrm{hom}$. Because $H^\mathrm{hom}$ is diagonal in Fourier space, $\mathbf{P}$ can be applied matrix-free using two FFTs and a pointwise multiplication, with overall complexity $\mathcal{O}(N\log N)$. The preconditioner $\mathbf{P}$ is defined as:
\begin{equation}
\label{eq: Vector_product_preconditioned}
P_{ij} v_j = - \frac{N}{2\pi} \sum_k \frac{1}{(1-\abs{k})\bar{G}(a_0)+a_0\bar{G}'(a_0)-\Gc^0} \hat{v}_k e^{+i k \theta_i}.
\end{equation}
i.e. it rescales each Fourier mode by the inverse homogeneous eigenvalue $\lambda_k$. This preconditioner gives speedups of a factor $2$ to $5$, depending on Poisson’s ratio $\nu$ and the heterogeneity contrast.

\section{Derivation of the analytical solutions for the validation cases}
\label{apdx:axisymetrical_solution}

In this appendix, we present first-order analytical solutions for coplanar crack propagation. We will focus on two cases: (1) a crack propagating in mode {II}+{III} (arbitrary $\nu$) in a homogeneous medium, and a crack propagating in mode~I or mode {II}+{III} ($\nu = 0$) in a heterogeneous fracture energy field $\Gc(r,\theta) = \Gc^0 \left(1 + \Delta \cos{p\theta}\right)$, where $p$ is the wavenumber of the perturbation, and $\Delta$ its amplitude. Using a linear superposition, we will retrieve a first order solution for the \textit{penny-shaped shear crack} in heterogeneous field with $\nu \neq 0$.

    \begin{remark}
    In Section~\ref{subsec:potential_energy}, we retained the full dependence on $\nu$ in Eq.~\ref{eq:Pi_pot_perturbed} even though our asymptotic expansion is only first-order in $\da$. While treating $\nu$ at all orders goes beyond strict first-order consistency \citep{favier_coplanar_2006}, it allows us to identify, by comparison with the strictly first-order analytical solutions derived in~\ref{apdx:axisymetrical_solution}, the range of $\nu$ over which our model remains quantitatively reliable, as shown in Section~\ref{subsec:validation}.
    \end{remark}

\subsection{Preliminaries}

We chose to restrict ourselves to the case where the crack is loaded by some system of arbitrary forces that generate an axisymmetric stress field $\sigma_{yy}(r)$ (in mode~I) and $\sigma_{yz}(r)$ (in mode~{II}+{III}) along the fracture plane.

For mode~I cracks, the mode~I SIF $\KIcirc$ along the front of a penny-shaped crack of radius $a_0$ reads:
\begin{align}
\label{eq:axisymmetric_SIF_1}
    \KIcirc(a_0) = \frac{2}{\pi}\sqrt{\pi a_0} \int_{0}^{1} \sigma_{yy}(a_0u) \dfrac{u}{\sqrt{1-u^2}}\,\mathrm{d}u.
\end{align}
For mode~{II}+{III} cracks, $\KIIcirc$ and $\KIIIcirc$ can be expressed as \citep{fabrikant_applications_1989, saez_three-dimensional_2022}:
\begin{align}
\label{eq:axisymmetric_SIF_2+3}
    \KIIcirc(a_0,\theta) =\left(+k_0(a_0) + \dfrac{\nu}{1-\nu/2} k_1(a_0) \right)\cos{\theta} \\
    \KIIIcirc(a_0,\theta) =\left(-k_0(a_0) + \dfrac{\nu}{1-\nu/2} k_1(a_0) \right) \sin{\theta}.
\end{align}
where:
\begin{align}
\label{eq:k0_and_k1}
    k_0(a_0) = \frac{2}{\pi}\sqrt{\pi a_0} \int_{0}^{1} \sigma_{yz}(a_0u) \dfrac{u}{\sqrt{1-u^2}}\,\mathrm{d}u \text{, and } k_1(a_0) = \frac{3}{\pi}\sqrt{\pi a_0} \int_{0}^{1} \sigma_{yz}(a_0u)u\sqrt{1-u^2}\,\mathrm{d}u.
\end{align}

For mode~{II}+{III}, the crack is loaded by a pair of shear point forces of intensity $Q$ in opposite directions, such that one has:
\begin{equation}
\label{eq:axisymmetric_point_force}
    k_0(a_0) = \dfrac{Q}{(\pi a_0)^{3/2}} \text{, and } k_1(a_0) = \frac{3}{2}k_0(a_0),
\end{equation}
which yields Eq.~\eqref{eq:SIFs_pointforce_Q}, while mode~I SIF is given in Eq.~\eqref{eq:SIFs_pointforce_P}.

\subsection{First-order solution for a shear crack propagating in a homogeneous material}

We consider first crack growth in homogeneous media of fracture energy $\Gc^0$. In mode~I, under axisymmetric loading conditions, the crack remains circular. In mode {II}+{III}, the energy release rate along the front of circular crack is not uniform, except when $\nu = 0$. The crack front must then deform to ensure that $G = \Gc$ everywhere along the front during propagation. 

We write:
\begin{equation}
\label{eq:homogeneous_a}
a(\nu,\theta) = a_0 + \nu a_1(\theta) + \mathcal{O}(\nu^2),
\end{equation}
and:
\begin{equation}
\label{eq:homogeneous_Kcirc}
\begin{aligned}
    \KIIcirc(a_0, \theta) & = \KIIcirc^\mathbf{(0)}(a_0, \theta) + \nu\KIIcirc^\mathbf{(1)}(a_0, \theta) + \mathcal{O}(\nu^2) = +k_0(a_0)\cos\theta + \nu k_1(a_0)\cos\theta + \mathcal{O}(\nu^2),\\
    \KIIIcirc(a_0, \theta) & = \KIIIcirc^\mathbf{(0)}(a_0, \theta) + \nu\KIIIcirc^\mathbf{(1)}(a_0, \theta) + \mathcal{O}(\nu^2) = -k_0(a_0)\sin\theta + \nu k_1(a_0)\sin\theta + \mathcal{O}(\nu^2),
\end{aligned}
\end{equation}
so that the first-order expansion of modes {II}/{III} SIFs along the perturbed crack front are expressed as:
\begin{equation}
\label{eq:homogeneous_K}
\begin{aligned}
    \KII(a_0, \theta) & = \KII^\mathbf{(0)}(a_0, \theta) + \nu\KII^\mathbf{(1)}(a_0, \theta) + \mathcal{O}(\nu^2) ,\\
    \KIII(a_0, \theta) & = \KIII^\mathbf{(0)}(a_0, \theta) + \nu\KIII^\mathbf{(1)}(a_0, \theta) + \mathcal{O}(\nu^2),
\end{aligned}
\end{equation}
where:
\begin{equation}
\label{eq:homogeneous_dK}
\begin{aligned}
    \KII^\mathbf{(0)}(a_0, \theta) & =  \KIIcirc^\mathbf{(0)}(a_0, \theta), \\
    \KIII^\mathbf{(0)}(a_0, \theta) & = \KIIIcirc^\mathbf{(0)}(a_0, \theta),  \\
    \KII^\mathbf{(1)}(a_0, \theta) & = \KIIcirc^\mathbf{(1)}(a_0, \theta) + \KIIpert^\mathbf{(\nu=0)}(a_0, [\KIIcirc^\mathbf{(0)}], [\KIIIcirc^\mathbf{(0)}], [a_1]; \theta)\\
    \KIII^\mathbf{(1)}(a_0, \theta) & = \KIIIcirc^\mathbf{(1)}(a_0, \theta) + \KIIIpert^\mathbf{(\nu=0)}(a_0, [\KIIcirc^\mathbf{(0)}], [\KIIIcirc^\mathbf{(0)}], [a_1]; \theta)
\end{aligned}
\end{equation}

Using Irwin's criterion of Eq.~\eqref{eq:Irwin_formula}, one gets for the first-order expansion of the energy release rate $G$:
\begin{equation}
\label{eq:homogeneous_G}
G(a_0, \theta) = G^\mathbf{(0)}(a_0, \theta) + \nu G^\mathbf{(1)}(a_0, \theta) + \mathcal{O}(\nu^2),
\end{equation}
where:
\begin{equation}
\label{eq:homogeneous_dG}
\begin{aligned}
    G^\mathbf{(0)}(a_0, \theta) & =  \dfrac{1}{E}\KII^\mathbf{(0)}(a_0, \theta)^2 + \dfrac{1}{E}\KIII^\mathbf{(0)}(a_0, \theta)^2, \\
    G^\mathbf{(1)}(a_0, \theta) & =  \dfrac{2}{E}\KII^\mathbf{(0)}(a_0, \theta)\KII^\mathbf{(1)}(a_0, \theta) + \dfrac{2}{E}\KIII^\mathbf{(0)}(a_0, \theta)\KIII^\mathbf{(1)}(a_0, \theta) + \dfrac{1}{E}\KIII^\mathbf{(0)}(a_0, \theta)^2.
\end{aligned}
\end{equation}

Injecting  Eq.~\eqref{eq:homogeneous_dG} into Griffith's criterion $G=\Gc^0$, one gets at zero order in $\nu$:
\begin{equation}
\label{eq:homogeneous_a0}
\dfrac{k_0(a_0)^2}{E} = \Gc^0,
\end{equation}
which gives $a_0$ as a function of the loading parameter (see Eq.~\eqref{eq:analytical_solution_heterogeneous} for point force loading).

At first order in $\nu$, one has:
\begin{equation}
\label{eq:homogeneous_G1}
G^\mathbf{(1)}(a_0, \theta) = \dfrac{2}{E}\KII^\mathbf{(0)}(a_0, \theta)\KII^\mathbf{(1)}(a_0, \theta) + \dfrac{2}{E}\KIII^\mathbf{(0)}(a_0, \theta)\KIII^\mathbf{(1)}(a_0, \theta) + \dfrac{1}{E}\KIII^\mathbf{(0)}(a_0, \theta)^2 = 0,
\end{equation}

Solving this system for $a_1$ gives: 
\begin{equation}
\label{eq:homogeneous_a1}
a(\theta, \nu) = a_0 \left(1 - \nu \frac{k_0(a_0)}{4a_0k'_0(a_0)} - \nu \cos(2\theta) \frac{k_0(a_0) - 4k_1(a_0)}{4(k_0(a_0) - a_0k'_0(a_0))}\right) + \mathcal{O}(\nu^2),
\end{equation}
Similar solutions, albeit less general, have been obtained by \cite{gao_nearly_1988} and \cite{favier_coplanar_2006}. However, our method provides a generic way to account for the non-isotropic crack growth in homogeneous materials. Changes in loading conditions (e.g. a gradient in shear stress in one direction) can also be superimposed through an additional perturbation of the reference SIFs $\KIIcirc$ and $\KIIIcirc$. Substituting Eq.~\eqref{eq:axisymmetric_point_force} in \eqref{eq:homogeneous_a1} yields the particular Eq.~\eqref{eq:analytical_solution_homogeneous} for point force loading.

\subsection{First-order solution for tensile and shear cracks propagating in a heterogeneous material}

 Here we focus on a specific heterogeneous field where the fracture energy is invariant in the radial direction $e_r$. This configuration generates stable crack propagation without instabilities. Namely, we use the fields of \cite{sanner_crack-front_2022} and consider ray-shaped fracture energy fields:
\begin{equation}
\label{eq:rays_Gc}
\Gc(\theta) = \Gc^0 \left( 1 + \Delta \cos{p \theta} \right),
\end{equation}
where $\Gc^0$ is the average fracture energy, $\Delta$ is the amplitude of the fracture energy perturbation and $p$ its wavenumber.

At first-order, contributions of the Poisson ratio $\nu$ and the heterogeneity amplitude $\Delta$ add up, so that it is possible to set $\nu=0$. General solutions will be obtained by adding up the $\nu$ contribution of Eq.~\eqref{eq:homogeneous_a1} with the $\Delta$ contribution that we will obtain in Eq.~\eqref{eq:rays_a1}.

We follow the same procedure as in the previous example, and expand the front position following:
\begin{equation}
\label{eq:rays_a}
a(\nu,\theta) = a_0 + \Delta a_1(\theta) + \mathcal{O}(\Delta^2),
\end{equation}
so that the first-order expansion modes {II}/{III} SIFs along the perturbed crack front are expressed as:
\begin{equation}
\label{eq:rays_K}
\begin{aligned}
    \KII(a_0, \theta) & = \KII^\mathbf{(0)}(a_0, \theta) + \Delta\KII^\mathbf{(1)}(a_0, \theta) + \mathcal{O}(\Delta^2) ,\\
    \KIII(a_0, \theta) & = \KIII^\mathbf{(0)}(a_0, \theta) + \Delta\KIII^\mathbf{(1)}(a_0, \theta) + \mathcal{O}(\Delta^2),
\end{aligned}
\end{equation}
where:
\begin{equation}
\label{eq:rays_dK}
\begin{aligned}
    \KII^\mathbf{(0)}(a_0, \theta) & =  \KIIcirc^\mathbf{(\nu=0)}(a_0, \theta), \\
    \KIII^\mathbf{(0)}(a_0, \theta) & = \KIIIcirc^\mathbf{(\nu=0)}(a_0, \theta),  \\
    \KII^\mathbf{(1)}(a_0, \theta) & = \KIIpert^\mathbf{(\nu=0)}(a_0, [\KIIcirc^\mathbf{(\nu=0)}], [\KIIIcirc^\mathbf{(\nu=0)}], [a_1]; \theta)\\
    \KIII^\mathbf{(1)}(a_0, \theta) & = \KIIIpert^\mathbf{(\nu=0)}(a_0, [\KIIcirc^\mathbf{(\nu=0)}], [\KIIIcirc^\mathbf{(\nu=0)}], [a_1]; \theta)
\end{aligned}
\end{equation}
The first-order expansion of the energy release rate $G$ thus reads:
\begin{equation}
\label{eq:rays_G}
G(a_0, \theta) = G^\mathbf{(0)}(a_0, \theta) + \Delta G^\mathbf{(1)}(a_0, \theta) + \mathcal{O}(\Delta^2),
\end{equation}
where:
\begin{equation}
\label{eq:rays_dG}
\begin{aligned}
    G^\mathbf{(0)}(a_0, \theta) & =  \dfrac{1}{E}\KII^\mathbf{(0)}(a_0, \theta)^2 + \dfrac{1}{E}\KIII^\mathbf{(0)}(a_0, \theta)^2, \\
    G^\mathbf{(1)}(a_0, \theta) & =  \dfrac{2}{E}\KII^\mathbf{(0)}(a_0, \theta)\KII^\mathbf{(1)}(a_0, \theta) + \dfrac{2}{E}\KIII^\mathbf{(0)}(a_0, \theta)\KIII^\mathbf{(1)}(a_0, \theta).
\end{aligned}
\end{equation}

Injecting Eq.~\eqref{eq:rays_dG} into Griffith's criterion $G=\Gc$, one finds at zero order in $\Delta$:
\begin{equation}
\label{eq:rays_a0}
\dfrac{k_0(a_0)^2}{E} = \Gc^0,
\end{equation}
which is equal to Eq.~\eqref{eq:homogeneous_a0} as the two problems are identical.

At first order in $\Delta$, one has:
\begin{equation}
\label{eq:rays_G1}
G^\mathbf{(1)}(a_0, \theta) = \dfrac{2}{E}\KII^\mathbf{(0)}(a_0, \theta)\KII^\mathbf{(1)}(a_0, \theta) + \dfrac{2}{E}\KIII^\mathbf{(0)}(a_0, \theta)\KIII^\mathbf{(1)}(a_0, \theta) = \Gc^0\cos{p\theta},
\end{equation}
which gives
\begin{equation}
\label{eq:rays_a1}
\begin{aligned}
& a(\theta)=a_0\left(1-\Delta \frac{1}{p-2 a_0 k_0^{\prime}\left(a_0\right) / k_0\left(a_0\right)} \cos p \theta\right) + \mathcal{O}(\Delta^2) & \text { if } p>1, \\
\text{ and } & a(\theta)=a_0\left(1-\Delta \frac{3}{-1 / 2-2 a_0 k_0^{\prime}\left(a_0\right) / k_0\left(a_0\right)} \cos \theta\right) + \mathcal{O}(\Delta^2) & \text { if } p=1,
\end{aligned}
\end{equation}
Substituting Eq.~\eqref{eq:axisymmetric_point_force} in Eq.~\eqref{eq:rays_a1}, and adding the $\nu$-contribution of Eq.~\eqref{eq:homogeneous_a1}, one finds Eq.~\eqref{eq:analytical_solution_heterogeneous} for point force loading. 

Similar reasoning for mode~I loading yields:
\begin{equation}
\label{eq:tensile_a1}
\begin{aligned}
& a(\theta)=a_0\left(1-\Delta \frac{1}{p-2 a_0 \KIcirc^{\prime}\left(a_0\right) / \KIcirc\left(a_0\right)} \cos p \theta\right) + \mathcal{O}(\Delta^2),
\end{aligned}
\end{equation}
where $a_0$ is solution of:
\begin{equation}
\label{eq:tensile_a0}
\dfrac{1-\nu^2}{E}\KIcirc(a_0)^2 = \Gc^0,
\end{equation}
which is valid no matter $\nu$.

\section{Convergence study}
\label{apdx:convergence_study}

We perform a convergence study to identify suitable values for the loading increment $\Delta a$ (prescribed via a uniform increase of the reference radius $a_0$), the number of front discretization points $N$, and the maximum trust-region radius $\Delta_\mathrm{tr}$. Simulations are run on disordered toughness fields with correlation length $d$, at the largest disorder level considered $\sigma/\Gc^0=0.5$ (i.e., $\Gc^\mathrm{max}/\Gc^\mathrm{min}\simeq 14$), and for $\nu=0.2$, which yields the largest front distortions and the strongest intermittency. In each test, we fix two parameters and vary the third, refining the discretization (i.e. decreasing $\Delta a$ or increasing $N$) or tightening the optimization constraint (i.e. decreasing $\Delta_\mathrm{tr}$).

For each parameter, we report three diagnostics that capture both accuracy and computational cost:
\begin{enumerate}
\item the crack-surface area variation $\Delta S$ relative to the homogeneous reference with toughness $\Gc^0$, defined in Eq.~\eqref{eq:area_variation}. As shown in Section~\ref{sec:applications}, $\Delta S$ is sensitive to both intermittency and apparent toughening;
\item the computation time, to quantify algorithmic performance;
\item the mean numerical error relative to the most refined case.
\end{enumerate}

\subsection{Step size $\Delta a$}

The step size $\Delta a$ sets the loading increments in our quasi-static scheme. Although the specimen is driven by point forces $P$ or $Q$, we parametrize loading by an associated increase of the reference radius $a_0$ of the homogeneous case. For a disorder field with a single correlation length $d$, $\Delta a$ measures how finely the loading samples the heterogeneity landscape: $\Delta a/d$ is the front advance per step in units of one heterogeneity.

\begin{figure}[h]
		\centering
		\noindent\includegraphics[width=0.9\textwidth]{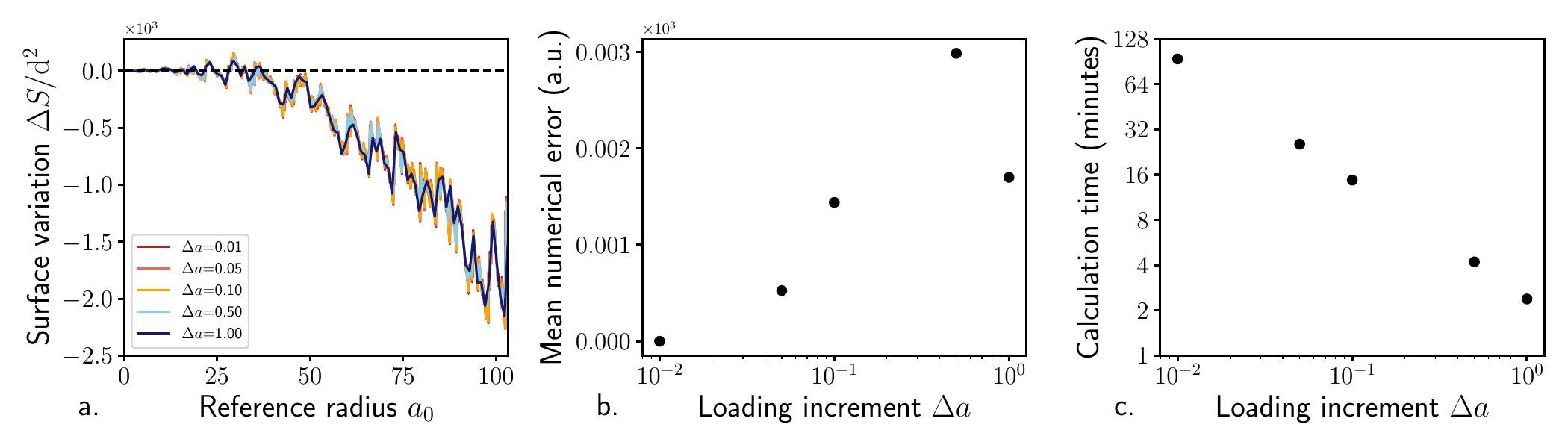}
		\caption{a. Surface area variation $\Delta S/d^2$ for increasing step size $\Delta a/d$ (in blue to red colored solid lines). b. Convergence error for increasing $\Delta a/d$ and c. calculation time.}
		\label{fig:convergence_study_increment}
	\end{figure}

We take $\Delta a \in [0.01\,d , d]$, $N=4096$, $\Delta_\mathrm{tr}=0.1\,d$. Figure~\ref{fig:convergence_study_increment}a shows that large step size smooth out intermittency. Importantly, the computed equilibrium front positions are the same for large and small $\Delta a$, indicating that the trust-region strategy prevents unphysical barrier crossing. For $\Delta a \lesssim 0.5\,d$ (at least two steps per heterogeneity), $\Delta S$ becomes visually insensitive to further refinement, while runtime decreases linearly with $\Delta a$. We therefore take $\Delta a=0.1\,d$ as a compromise between accuracy and total runtime. Consistently, Fig.~\ref{fig:convergence_study_increment}b indicates robust convergence of the minimization across the tested range of $\Delta a$. Runtime scales approximately linearly with $\Delta a$ (Fig.~\ref{fig:convergence_study_increment}c).

\subsection{Number of discretization points $N$}

While the step size $\Delta a$ sets the radial resolution, the number of discretization points $N$ sets the resolution in the orthoradial direction. Because $N$ is held fixed during propagation, the number of points per heterogeneity $\simeq N/(2\pi \left<a\right>)$ progressively drops as the mean radius $\left<a\right>$ grows. At the largest crack sizes $\left<a\right> \simeq 100\, d$ the front can under-sample weak/tough patches. This loss of resolution affects the front roughness, the onset of instabilities and crack arrest.

\begin{figure}[h]
		\centering
		\noindent\includegraphics[width=0.9\textwidth]{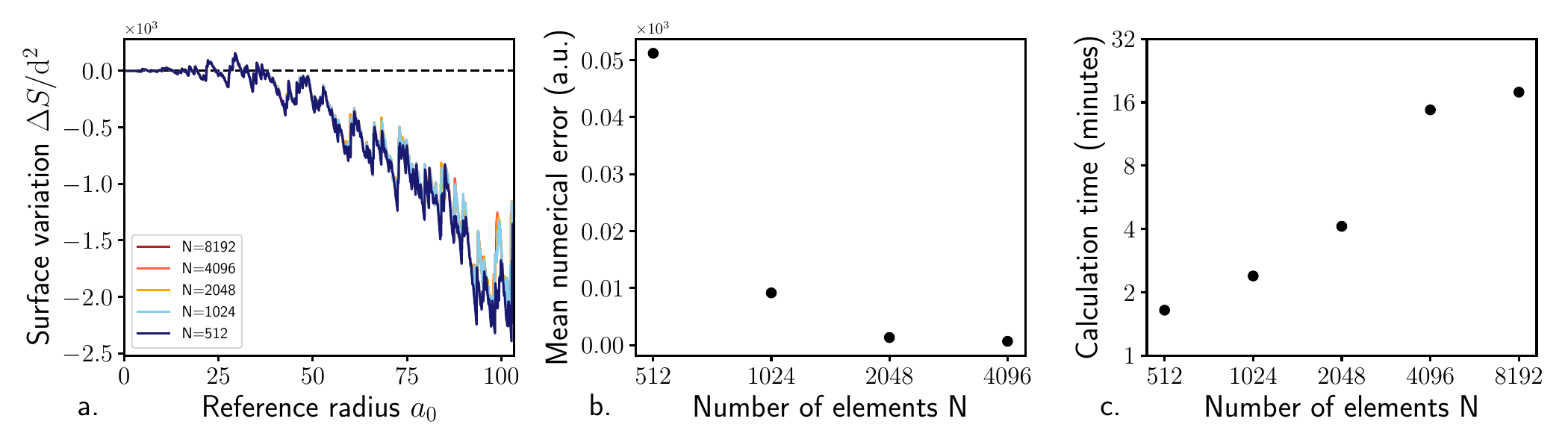}
		\caption{a. Normalized surface variation $\Delta S/d^2$ calculated with increasing number $N$ of discretization points (in blue to red colored solid lines). b. Convergence error for increasing $N$ and c. calculation time.}
		\label{fig:convergence_study_elements}
	\end{figure}

We take $N \in [512, 8192]$, $\Delta a = 0.1\,d$, $\Delta_\mathrm{tr}=0.1\,d$. Figure~\ref{fig:convergence_study_elements}a-b shows that coarse discretizations degrade the accuracy of our algorithm, with stable configurations shifted towards smaller $a$ values. As a result, underdiscretizing the fracture energy field leads to an increase of apparent toughening. Runtime scales approximately linearly with $N$ (Fig.~\ref{fig:convergence_study_elements}c). We choose $N=2048$ for the statistical campaign of 116,000 simulations.

\subsection{Maximum trust region radius} 

\begin{figure}[H]
		\centering
		\noindent\includegraphics[width=0.9\textwidth]{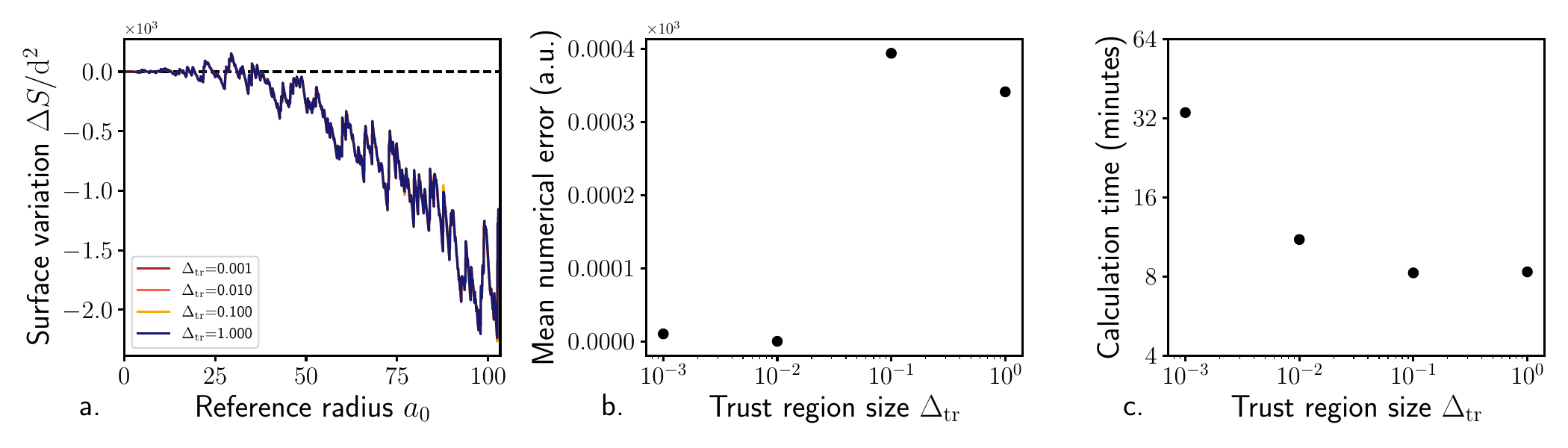}
		\caption{a. Surface area variation $\Delta S/d^2$ for increasing maximum trust region radius $\Delta_\mathrm{tr}/d$ (in blue to red colored solid lines). b. Convergence error for increasing $\Delta_\mathrm{tr}/d$ and c. calculation time.}
		\label{fig:convergence_study_trust}
	\end{figure}

The maximum trust-region radius $\Delta_\mathrm{tr}$ prevents the solver from converging to a distant local minimum instead of the next physically reachable equilibrium. We take $\Delta_\mathrm{tr}\in[10^{-3}d,10d]$ with $N=4096$ and $\Delta a=0.1,d$. Figure~\ref{fig:convergence_study_trust}a-b shows that $\Delta S$ is essentially insensitive to $\Delta_\mathrm{tr}$ over $10^{-3}d$ to $d$. Note that the algorithm did not converge for $\Delta_\mathrm{tr}=10\, d$. Runtime increases only for $\Delta_\mathrm{tr}=10^{-3}d$ (Fig.~\ref{fig:convergence_study_trust}c). As a result, we fix $\Delta_\mathrm{tr}=10^{-1}d$ in the remainder of the study.

\bibliographystyle{elsarticle-harv}
\bibliography{article_main}
	
\end{document}